\documentclass[journal,draftcls,onecolumn,12pt, double]{IEEEtran}
\usepackage{subfigure,cite,graphicx,psfig,amsmath,amssymb,eufrak,mathrsfs,epsf,epsfig}
\usepackage[usenames]{color}
\usepackage{amsmath}
\usepackage{multicol,lipsum}
\usepackage{xfrac}

\usepackage[active]{srcltx}
\ifCLASSINFOpdf
\else
\fi

\begin{document}
\newtheorem{ach}{Achievability}
\newtheorem{con}{Converse}
\newtheorem{definition}{Definition}%[section]
\newtheorem{theorem}{Theorem}%[section]
\newtheorem{lemma}{Lemma}%[section]
\newtheorem{example}{Example}
\newtheorem{cor}{Corollary}%[section]
\newtheorem{prop}{Proposition}%[section]
\newtheorem{conjecture}{Conjecture}%[section]
\newtheorem{remark}{Remark}%[section]

\def \tr{\operatorname{tr}}
\def \rank{\operatorname{rank}}

% can use \linebreak \\ within to get better formatting as desired
\title{Integer Forcing Interference Management\\ for the MIMO Interference Channel }\author{\IEEEauthorblockN{Sung Ho Chae,~\IEEEmembership{Member,~IEEE} and Sang-Woon Jeon,~\IEEEmembership{Member,~IEEE}
\thanks{S. H. Chae is with the Department of Electronic Engineering, Kwangwoon University, Seoul 01897, South Korea (e-mail: sho.chae00@gmail.com).}
\thanks{S.-W. Jeon, \emph{corresponding author}, is with the Department of Military Information Engineering, Hanyang University, Ansan 15588, South Korea (e-mail: sangwoonjeon@hanyang.ac.kr).}
}}
 \maketitle

\begin{abstract}
A new interference management scheme based on integer forcing (IF) receivers is studied for the two-user multiple-input and multiple-output (MIMO) interference channel. The proposed scheme employs a message splitting method that divides each data stream into common and private sub-streams, in which the private stream is recovered by the dedicated receiver only while the common stream is required to be recovered by both receivers. Specifically, to enable IF sum decoding at the receiver side, all streams are encoded using the same lattice code. Additionally, the number of common and private streams of each user is carefully determined by considering the number of antennas at transmitters and receivers, the channel matrices, and the effective signal-to-noise ratio (SNR) at each receiver to maximize the achievable rate. Furthermore, we consider various assumptions of channel state information at the transmitter side (CSIT)  and propose low-complexity linear transmit beamforming suitable for each CSIT assumption. The achievable sum rate and rate region are analytically derived and extensively evaluated by simulation for various environments, demonstrating that the proposed interference management scheme strictly outperforms the previous benchmark schemes in a wide range of channel parameters due to the gain from IF sum decoding.
\end{abstract}
\begin{IEEEkeywords}
Interference channel, interference management, integer forcing (IF), linear beamforming, multiple-input multiple-output (MIMO). 
\end{IEEEkeywords}
 \IEEEpeerreviewmaketitle

 %This observation is in contrast to the point-to-point channel, multiple access channel, and broadcast channel cases in which hybrid beam-forming cannot increase the sum DoF

% except when all the interference can be eliminated without the need to use more antennas
% when interferences among users cannot be entirely eliminated
\section{Introduction}
In the fifth generation (5G) and the future sixth generation (6G) communication systems, it is expected that a variety of emerging applications such as virtual reality
(VR), augmented reality (AR), real-time ultra high definition (UHD) video streaming, autonomous driving, and the Internet of everything (IoE) \textcolor{black}{will begin to be utilized in earnest}~\cite{Samsung:20, Dang:20,Giordani:20,ITU:2017,Jiang:21}. To provide these new services any time, anywhere, regardless of the user's location, it is important to significantly increase the achievable data rate of cell edge users compared to the current level. More specifically, the target data rates for edge users (bottom $5\%$ of users) are $100$ Mbps and $50$ Mbps for 5G downlink and uplink, respectively, and moreover, 
it is predicted that in 6G communication, it will be necessary to achieve data rates that are 10 times higher than at present~\cite{Samsung:20,Jiang:21,ITU:2017}. To achieve such challenging goal, unprecedented technology is required to push the limits of current technologies, and advanced interference management based on multiple-input multiple-output (MIMO) antennas is being considered as one of the key technologies to significantly improve data rates of edge users in  5G and 6G communications~\cite{Chae:16,Jiang:21,Chen:20,Zhang:19,Siddiqui:21,Saqib:21}.

\subsection{Related Works}
Currently, linear MIMO receivers such as zero-forcing (ZF) receivers and minimum mean square error (MMSE) receivers have been widely used in practice~\cite{Tse_wireless,Shafi:17,Nadeem:19} due to their low complexity. These conventional linear receivers first separate the transmitted streams by applying a linear filter to the received signal vector and then decode each stream individually. Specifically, a ZF receiver uses the pseudo-inverse of the channel matrix to convert a given MIMO channel into interference-free parallel single-input and single-output (SISO) channels while an MMSE receiver uses the regularized channel inversion matrix to maximize the signal-to-noise ratio (SNR) of each individual stream. It is well known that MMSE receivers outperform ZF receivers regardless of the SNR regime, and the performance gap becomes larger at lower SNR~\cite{Tse_wireless}. In addition, an MMSE receiver can be combined with successive interference cancellation (SIC)  operation to achieve better performance, called an MMSE-SIC receiver~\cite{Tse_wireless}. The MMSE-SIC receiver sequentially recovers each stream one by one by applying a linear MMSE filter, and in each sequential decoding, the contribution of the decoded stream is subtracted from the received signal vector to increase the SNR of the remaining streams that have not yet been decoded. Furthermore, ZF, MMSE, and MMSE-SIC receivers can be employed for intra-cell and/or inter-cell interference management in multi-user communications as well as point-to-point communications~\cite{Shafi:17,Nadeem:19}. 

Although the aforementioned linear receivers clearly have an advantage in terms of complexity, they have an inherent limitation that decoupling with linear filtering can cause significant noise amplification, and as a result, the gap between the achievable rate and the capacity can be large,  especially when the channel becomes near singular~\cite{Tse_wireless,Costello:07,Kumar:09,Jiang:11}. On the other hand, nonlinear receivers such as maximum likelihood (ML) detection receiver and sphere decoding receiver~\cite{Damen:03,Hassibi:05,Jalden:05}, which exhaustively search for the most likely transmitted signal vector based on the received signal vector, can significantly reduce  noise amplification compared to linear receivers. However, the complexity of most nonlinear receivers is  much higher than that of linear receivers especially when supporting large numbers of antennas and/or high order modulations~\cite{Guo:06}, and hence nonlinear receivers are still rarely used in practical communication systems~\cite{Shafi:17,Nadeem:19}.

Recently, a novel linear MIMO receiver, namely, \emph{integer forcing (IF) receiver}, has been proposed~\cite{Zhan:14}.  Note that in the previous work~\cite{Nazer:11}, the \emph{compute-and-forward (CF)} relaying scheme has been developed for Gaussian relay networks, in which the integer-linear sum of transmitted codewords is first computed at each relay and then the result is forwarded to the destination. The IF MIMO receiver can be viewed as a receiver to which the CF scheme is applied in a Gaussian MIMO point-to-point channel. Unlike the previous linear receivers that first separate the transmitted streams by applying a linear filter and recover each stream individually with each SISO decoder, the IF receiver first creates an integer-valued full-rank effective channel by applying a linear filter and directly decodes the integer-linear combination of the transmitted codewords with each SISO decoder. To enable this sum decoding, the integer-linear sum of the transmitted codewords should be itself a codeword, and hence, the transmitter uses the same lattice code for all streams. After integer-linear sums are decoded, the original streams can be simply recovered by  multiplying the decoded outputs of the linear combinations by the inverse of the effective integer channel matrix in the absence of noise. Since the IF receiver has the freedom to determine the effective integer channel matrix  in a way that minimizes noise amplification in contrast to the previous linear receivers that always constrain the integer matrix by the identity matrix regardless of the channel matrix, IF receivers can significantly reduce noise amplification compared to the previous linear receivers. Consequently, \textcolor{black}{it has been shown that IF receivers can achieve rate close to the channel capacity in Rayleigh fading channels}~\cite{Zhan:14}.

In the literature, several follow-up studies have been conducted to improve and extend the basic IF proposed in~\cite{Zhan:14}. In~\cite{Ordentlich:15}, it has been shown that using a linear  dispersion space-time code in conjunction with IF equalization at the receiver side can achieve the capacity within a constant gap for general MIMO channels, assuming that the transmitter only knows white-input mutual information. More recently, precise performance characteristics of parallel MIMO channels have been studied when applying precoding with the full-diversity rotation matrix at the transmitter side and IF equalization at the receiver side~\cite{Regev:21}. In addition, similar to MMSE-SIC, SIC operations can be combined with IF sum decoding, namely, \emph{successive IF}~\cite{Ordentlich:conf_13},  to improve the performance of the basic IF. 
Additionally, similar to ZF and MMSE receivers, IF receivers can be employed to manage interference in multi-user communications as well as point-to-point communications, i.e., it can be extended to multiple-access channels~\cite{He:18,Ordentlich:conf_13}, broadcast channels~\cite{Silva:17,He:18,Ahn:19,Venturelli:20}, interference channels~\cite{Ntranos:13,Hong:13,Ordentlich:14,Azimi-Abarghouyi:18}, and relay networks~\cite{Abarghouyi:16,Jiang:19,Hejazi:16}. \textcolor{black}{For example, in~\cite{Hejazi:16}, extended CF and successive CF have been proposed for multi-user multi-relay networks, and it has been shown that the proposed schemes can solve the rank failure problem and outperform the original CF scheme of~\cite{Nazer:11}.} Furthermore, instead of using lattice codes, IF  transceivers built on off-the-shelf binary codes such as turbo and low-density parity-check (LDPC) codes have been recently developed and evaluated in practical channel environments~\cite{Chae:16,Ahn-Chae-Kim-Kim:21}. In~\cite{Chae-Jeon-Ahn:19}, spatially modulated IF (SM-IF) that combines generalized spatial modulation with IF has been developed based on practical binary codes. 

\subsection{Our Contributions}
In this paper, we propose a low-complexity interference management scheme based on IF for the two-user MIMO interference channel. It is well known that when interference is strong, decoding interference can enlarge the achievable rate region~\cite{Costa87,Carleial75,Sato81,LNIT}. On the other hand, when interference is weak enough, treating interference as noise has proven to be an optimal strategy to achieve the capacity region~\cite{Annapureddy:09,Motahari09,Shang09}. Inspired by these facts, we consider a message splitting method that splits each data stream into common and private sub-streams, as in the Han--Kobayashi scheme~\cite{Han:81}. Each receiver then attempts to recover the desired streams, that is, the intended common and private streams, and also the other user's common streams, while treating the private streams of the other user as noise. The main difference between our work and  previous message splitting schemes is that unlike  previous studies, all common and private streams are encoded with the same lattice code to enable IF sum decoding at the receiver side in this paper. \textcolor{black}{Note that a lattice based message splitting scheme for IF receivers has been studied in~\cite{Ordentlich:14} as in our work,  but in the earlier work~\cite{Ordentlich:14}, the SISO interference channel was considered rather than the MIMO interference channel. Our work can be viewed as a follow-up study extending the results of~\cite{Ordentlich:14} to  MIMO networks.}

In the proposed scheme, novel interference management and rank adaptation based on IF sum decoding are performed  by controlling  the number of common and private streams for each user. For a given channel realization and SNR, the optimal choice that maximizes the achievable rate of the proposed scheme can be found in a finite search space that only depends on the number of antennas at transmitters and receivers. In addition, we propose low-complexity linear transmit beamforming  suitable for various channel state information at the transmitter side (CSIT) assumptions. The achievable sum rate and rate region of the proposed scheme are analytically derived and also numerically evaluated for various channel environments. 
The results demonstrate that the proposed scheme significantly improves the achievable rate compared to conventional previous schemes for a wide range of channel parameters. To the best of our knowledge, our results are the first to demonstrate such improvement by combining IF sum decoding  with a message-splitting method for the MIMO interference channel.

\subsection{Paper Organization and Notation} 
The remainder of the paper is organized as follows. In Section II, we describe the channel model and assumptions considered in this paper. In Section III, the proposed  interference management scheme based on IF is specifically described and the resulting achievable rate is derived. In Section IV, we numerically evaluate the achievable sum rate and rate region for various environments and discuss the results. Finally, we conclude the paper in Section V. 

\emph{Notation}: Boldface lowercase and uppercase letters are used to denote vectors and matrices, respectively. Let $\mathbf{A}^\dagger$, $\mathbf{A}^{-1}$, $\|\mathbf{A}\|$, and $\text{rank}(\mathbf{A})$ denote the transpose, the inverse, the norm, and the rank of $\mathbf{A}$, respectively. We denote the $n\times n$ identity matrix by $\mathbf{I}_n$.  The real and imaginary parts of $\mathbf{A}$ are denoted by $\textrm{Re}(\mathbf{A})$ and $\textrm{Im}(\mathbf{A})$, respectively. Let $[\mathbf{A}]_{i:j}$ denote the submatrix of $\mathbf{A}$ consisting of the $i$th to $j$th column vectors of  $\mathbf{A}$. Let us denote $\log^+(x)=\max\{\log_2(x),0\}$ and $[1:n]=\{1,2,\cdots,n\}$.  
We denote the circularly symmetric complex Gaussian distribution with mean $0$ and variance $\sigma^2$ by $\mathcal{CN}(0, \sigma^2)$ and the continuous uniform distribution on the interval $[a,b]$ by $\text{Unif} [a,b]$. The expectation is denoted by $\mathbb{E}(\cdot)$.
%
%\begin{figure}[t!]
%\begin{center}
%\includegraphics[scale=0.7]{System_model.eps}
%\end{center}
%\vspace{-0.1in}
%\caption{The $K$-user interference channel with energy harvesting}
%\label{system}
%\vspace{-0.1in}
%\end{figure}

\section{System Model}
Consider the two-user MIMO interference channel, in which  transmitter $i$ attempts to communicate with receiver $i$ while interfering to receiver $j\neq i$, where $i,j\in\{1,2\}$. Each transmitter and receiver are equipped with $M_\text{T}$ antennas and $M_\text{R}$ antennas, respectively. 
It is assumed that communication takes place over $n$ time slots. Then the received signal vector of receiver $i$ at time slot $t$, denoted by $\mathbf{y}_i(t)\in \mathbb{C}^{M_\text{R}\times 1}$, is given by
\begin{align}\label{eq:system model}
\mathbf{y}_i(t)=\sum_{j=1}^2\mathbf{H}_{i,j}(t)\mathbf{x}_j(t)+\mathbf{z}_i(t),~t\in[1:n],
\end{align}
where $\mathbf{x}_j(t) \in \mathbb{C}^{M_\text{T} \times 1}$ is the complex-valued transmit signal vector from transmitter $j$ at time slot $t$,  $\mathbf{H}_{i,j} (t) \in \mathbb{C}^{M_\text{R} \times M_\text{T}}$ is the complex-valued channel matrix from transmitter $j$ to receiver $i$ at time slot $t$, and $\mathbf{z}_i(t)$ is the complex additive white Gaussian noise vector of receiver $i$ at time slot $t$ with  $\mathbf{z}_i(t)\sim \mathcal{CN}\left(\mathbf{0}_{M_\text{R}\times M_\text{R}}, \mathbf{I}_{M_\text{R}}\right)$. Note that the complex-valued input--output relation~\eqref{eq:system model} can be equivalently represented in the form of a real-valued expression as
\begin{align}\label{eq:real-valued channel model}
\bar{\mathbf{y}}_i(t)=\sum_{j=1}^2\bar{\mathbf{H}}_{i,j}(t)\bar{\mathbf{x}}_j(t)+\bar{\mathbf{z}}_i(t),~t\in[1:n],
\end{align}
where 
\begin{align*}
\bar{\mathbf{y}}_i(t)&=\left[\begin{array}{cc}\textrm{Re}(\mathbf{y}_i(t))\\ \textrm{Im}(\mathbf{y}_i(t))\end{array}\right]\in\mathbb{R}^{2M_{\text R}\times 1},\\
\bar{\mathbf{H}}_{i,j}(t)&=\left[\begin{array}{cc}\textrm{Re}(\mathbf{H}_{i,j}(t))& -\textrm{Im}(\mathbf{H}_{i,j}(t)) \\ \textrm{Im}(\mathbf{H}_{i,j}(t)) & ~~\textrm{Re}(\mathbf{H}_{i,j}(t))\end{array}\right]\in \mathbb{R}^{2M_{\text R}\times  2M_{\text T}},\\
\bar{\mathbf{x}}_j(t)&=\left[\begin{array}{cc}\textrm{Re}(\mathbf{x}_j(t))\\ \textrm{Im}(\mathbf{x}_j(t))\end{array}\right]\in\mathbb{R}^{2M_{\text T}\times 1},\\
\bar{\mathbf{z}}_i(t)&=\left[\begin{array}{cc}\textrm{Re}(\mathbf{z}_i(t))\\ \textrm{Im}(\mathbf{z}_i(t))\end{array}\right]\in\mathbb{R}^{2M_{\text R}\times 1}.
\end{align*}
For notational convenience, we consider the real-valued channel stated in~\eqref{eq:real-valued channel model} hereafter. In addition, each transmitter should satisfy the average transmit power constraint $P$, i.e., $\frac{1}{n}\sum_{t=1}^{n}\|\bar{\mathbf{x}}_i(t)\|^2\leq P$,  $\forall i\in \{1,2\}$.

We assume that channel matrices are static during communication, i.e., $\bar{\mathbf{H}}_{i,j}(t)=\bar{\mathbf{H}}_{i,j}$, $\forall t\in[1:n]$, and thus the time index $t$ is omitted from the notation of channel matrices hereafter. In addition, we assume global channel state information  at the receiver side (CSIR), i.e., $\bar{\mathbf{H}}_{i,j}$ is known to each receiver for all $i.j\in\{1,2\}$. On the other hand, regarding CSIT, we consider three different scenarios: 1) no CSI is available at the transmitter side, i.e., all channel matrices are unknown to both transmitters; 2) only the CSI of its own desired link is available at each transmitter, i.e., 
transmitter $i\in\{1,2\}$ knows the coefficients of the desired channel matrix $\bar{\mathbf{H}}_{i,i}$ but does not know those of the other channel matrices; 3) 
each transmitter knows the CSI of both its desired link and interfering link, i.e., the coefficients in $\bar{\mathbf{H}}_{1,i}$ and $\bar{\mathbf{H}}_{2,i}$ are known to transmitter $i\in\{1,2\}$. In the next section, we will propose a novel transmission scheme using IF that adequately mitigates interference  with low complexity  for each of the aforementioned CSI assumptions.

%where $\mathbf{x}_i(t),\mathbf{x}_j(t) \in \mathbb{C}^{M_\text{R} \times 1}$ are the input vectors of transmitters $i$ and $j$, respectively, where $i\neq j$,
% $\mathbf{H}_{i,i}(t),\mathbf{H}_{i,j}(t), \in \mathbb{C}^{M_\text{R} \times 1}$ 

\section{Proposed IF-based Transmission Scheme}

\subsection{Transmitter Side}
\subsubsection{Nested lattice codes} We adopt a lattice coding scheme for IF in~\cite{Zhan:14,Ordentlich:conf_13,Ordentlich:14,Ordentlich:15,Nazer:11,Regev:21} and extend it in a way suitable for the two-user MIMO interference channel. 
For completeness, we briefly review the codebook construction of nested lattice codes here and refer to~\cite{Zamir_book} for more detailed definitions and properties of lattice codes.

Consider an $n$-dimensional lattice $\Lambda\in \mathbb{R}^n$, which is a discrete subgroup of $\mathbb{R}^n$ closed to any integer-linear combinations. 
The modulo-$\Lambda$ operation on a length-$n$ vector $\mathbf{c}$ is defined as 
\begin{align}
\mathbf{c}~\text{mod}~\Lambda=\mathbf{c}-Q_\Lambda(\mathbf{c}), 
\end{align}
where $Q_\Lambda(\mathbf{c})$ is the nearest neighbor quantizer that calculates 
\begin{align}
Q_\Lambda(\mathbf{c})= \underset{\mathbf{t}\in \Lambda}{\mathrm{argmin}} \|\mathbf{c}-\mathbf{t}\|.
\end{align}
The Voronoi region of $\Lambda$ is defined as $\mathcal{V}_{\Lambda}=\{\mathbf{c}\in \mathbb{R}^n : Q_{\Lambda}(\mathbf{c}
)=\mathbf{0}\}$. Additionally, the second moment of $\Lambda$ per dimension is defined as 
\begin{align}\label{second moment}
P_\Lambda=\frac{1}{n}\frac{1}{\text{Vol}(\mathcal{V}_\Lambda)}\int_{\mathbf{c}\in\mathcal{V}_\Lambda}\|\mathbf{c}\|^2 d\mathbf{c}, 
\end{align}
where $\text{Vol}(\mathcal{V}_\Lambda)$ denotes the volume of $\mathcal{V}_\Lambda$.

Now consider $n$-dimensional lattices $\Lambda_c$ and $\Lambda_f$. If $\Lambda_c \subseteq \Lambda_f$, then $\Lambda_c$ is said to be nested in  $\Lambda_f$, where $\Lambda_c$ and  $\Lambda_f$ are referred to as the coarse and fine lattices, respectively.
From the nested pair $\Lambda_c$ and $\Lambda_f$, a nested lattice codebook $\mathcal{C}$ can be constructed as $\mathcal{C}=\Lambda_f \cap\mathcal{V}_{\Lambda_c} $, where the associated code rate is given by 
\begin{align}
R=\frac{1}{n}\log \left|\mathcal{C} \right|=\frac{1}{n}\log \left|\Lambda_f \cap\mathcal{V}_{\Lambda_c} \right| ~\text{(bits/channel use)}.
\end{align}
Finally, since the transmit power constraint is given by $P$, the codebook $\mathcal{C}$ should be scaled to satisfy $P_{\Lambda_c}=P$, where $P_{\Lambda_c}$ can be obtained by substituting  $\mathcal{V}_{\Lambda_c}$ for $\mathcal{V}_\Lambda$  in~\eqref{second moment}.

\subsubsection{Encoding}\label{encoding}

For a non-negative integer $d_i$ satisfying that $d_i\leq \min \{2M_{\text{T}}, 2M_{\text{R}}\}$, let $\mathbf{w}_{i}\in \mathbb{Z}_2^{1\times d_i nR}$  denote the binary data stream for transmitter $i$ of length $d_i nR$, which is assumed to be independently and uniformly drawn from $\mathbb{Z}_2^{1\times d_i nR}$. 
To send $\mathbf{w}_{i}$ using multiple transmit antennas, it is equally partitioned into $d_{[\text{c}],i}+d_{[\text{p}],i}$ sub-streams of length $nR$ each. Specifically, denote 
\begin{align}
\mathbf{w}_{i}=\left[\begin{array}{ccccccccccccccccccccccc} \mathbf{w}_{[\text{c}],i,1} & \mathbf{w}_{[\text{c}],i,2}&\ldots& \mathbf{w}_{[\text{c}],i,d_{[\text{c}],i}}& \mathbf{w}_{[\text{p}],i,1}&\mathbf{w}_{[\text{p}],i,2}&\ldots& \mathbf{w}_{[\text{p}],i,d_{[\text{p}],i}}\end{array}\right], 
\end{align}
where  $\mathbf{w}_{[\text{c}],i,j}\in \mathbb{Z}_2^{1\times nR}$ and  $\mathbf{w}_{[\text{p}],i,k}\in \mathbb{Z}_2^{1\times nR}$ for all $j\in[1:d_{[\text{c}],i}]$, $k\in[1:d_{[\text{p}],i}]$, and  $i\in\{1,2\}$.
That is, $d_i=d_{[\text{c}],i}+d_{[\text{p}],i}\leq \min \{2M_{\text{T}}, 2M_{\text{R}}\}$.
Note that we here split $\mathbf{w}_{i}$ into common and private streams. Specifically, $\{\mathbf{w}_{[\text{p}],i,j}\}_{j\in[1:d_{[\text{p}],i}] }$ is the set of private streams intended to be recovered by receiver $i$ only, while $\{\mathbf{w}_{[\text{c}],i,k}\}_{i\in\{1,2\},k\in[1:d_{[\text{c}],i}]} $ is the set of common streams required to be recovered by both receivers $1$ and $2$, although streams in $\{\mathbf{w}_{[\text{c}],j,k}\}_{k\in[1:d_{[\text{c}],j}]} $ are not the intended streams for receiver $i\neq j$.

% Let $\mathbf{w}_{i}\in \mathbb{Z}_2^{1\times d_i nR}$  denote the binary data stream for transmitter $i$ of length $d_i nR$, where $i\in \{1:2\}$ and $d_i\leq \min \{2M_{\text{T}}, 2M_{\text{R}}\}$. It is assumed that each stream is independently and uniformly drawn from $\mathbb{Z}_2^{1\times d_i nR}$. In addition, $\mathbf{w}_{i}$ is divided into sub-streams $\mathbf{w}_{[\text{c}],i}\in \mathbb{Z}_p^{1\times d_{[\text{c}],i}nR}$ and $\mathbf{w}_{[\text{p}],i}\in \mathbb{Z}_p^{1\times d_{[\text{p}],i}nR}$, where $\mathbf{w}_i=[\mathbf{w}_{[\text{c}],i}, \mathbf{w}_{[\text{p}],i}]$ and $d_i=d_{[\text{c}],i}+d_{[\text{p}],i}$, $\forall i=1,2$. Moreover, $\mathbf{w}_{[\text{c}],i}$ and $\mathbf{w}_{[\text{p}],i}$ are  further partitioned into sub-streams $\{\mathbf{w}_{[\text{c}],i,1},\mathbf{w}_{[\text{c}],i,2},\ldots, \mathbf{w}_{[\text{c}],i,d_{[\text{c}],i}}\}$ and  $\{\mathbf{w}_{[\text{p}],i,1},\mathbf{w}_{[\text{p}],i,2},\ldots, \mathbf{w}_{[\text{p}],i,d_{[\text{p}],i}}\}$, respectively, $\forall i=1,2$, in which each of them has 

Then $\mathbf{w}_{[\text{c}],i,j}$ and  $\mathbf{w}_{[\text{p}],i,k}$ are encoded by the nested lattice code $\mathcal{C}\subset \mathbb{R}^{1\times n}$ explained above, i.e.,  $\mathbf{w}_{[\text{c}],i,j}$ and  $\mathbf{w}_{[\text{p}],i,k}$  are mapped into lattice points $\mathbf{b}_{[\text{c}],i,j}\in\mathcal{C}$ and $\mathbf{b}_{[\text{p}],i,k}\in\mathcal{C}$, respectively, for all $j\in[1:d_{[\text{c}],i}]$, $k\in[1:d_{[\text{p}],i}]$, and  $i\in\{1,2\}$. Notice that the same lattice code is assumed to be used for all streams $\{\mathbf{w}_{[\text{c}],i,j}\}_{i\in\{1,2\},j\in[1:d_{[\text{c}],i}] }$ and $\{\mathbf{w}_{[\text{p}],i,k}\}_{i\in\{1,2\},k\in[1:d_{[\text{p}],i}] }$ to enable IF sum decoding at each receiver. 

Next, a random dither $\mathbf{d}_{[\text{c}],i,j}\in\mathbb{R}^{1\times n}$, which is uniformly distributed over $\mathcal{V}_{\Lambda_c}$ and independent of $\mathbf{b}_{[\text{c}],i,j}$, is employed to generate $\mathbf{s}_{[\text{c}],i,j}=(\mathbf{b}_{[\text{c}],i,j}-\mathbf{d}_{[\text{c}],i,j})\mod \Lambda_c$, $\forall i\in\{1,2\}$ and $\forall j\in[1:d_{[\text{c}],i}]$. In the same manner, we can obtain $\mathbf{s}_{[\text{p}],i,k}=(\mathbf{b}_{[\text{p}],i,k}-\mathbf{d}_{[\text{p}],i,k})\mod \Lambda_c$, $\forall i\in\{1,2\}$ and $\forall k\in[1:d_{[\text{p}],i}]$, where $\mathbf{d}_{[\text{p}],i,k}\in\mathbb{R}^{1\times n}$ is a random dither  uniformly distributed over $\mathcal{V}_{\Lambda_c}$. It is assumed that all random dithers are known to all transmitters and receivers before communication. Note that due to the Crypto Lemma~\cite[Lemma 1]{Erez:04}, $\mathbf{s}_{[\text{c}],i,j}\in\mathbb{R}^{1\times n} $ and $\mathbf{s}_{[\text{p}],i,k}\in\mathbb{R}^{1\times n}$ are uniformly distributed over $\mathcal{V}_{\Lambda_c}$ and independent of $\mathbf{b}_{[\text{c}],i,j}$ and $\mathbf{b}_{[\text{p}],i,j}$, and  their variances are given by $\frac{1}{n}\mathbb{E}(\|\mathbf{s}_{[\text{p}],i,k}\|^2)=\frac{1}{n}\mathbb{E}(\|\mathbf{s}_{[\text{c}],i,k}\|^2)=P_{\Lambda_c}=P$.

Consequently, $\mathbf{s}_{[\text{c}],i,j}$ and $\mathbf{s}_{[\text{p}],i,k}$ are used as input signals to send streams  $\mathbf{w}_{[\text{c}],i,j}$ and  $\mathbf{w}_{[\text{p}],i,k}$, respectively. Since there are $d_{[\text{c}],i}+d_{[\text{p}],i}$ independent data streams with equal rate $R$ for transmitter $i\in\{1,2 \}$, the total transmission rate of transmitter $i$ is $R_i=(d_{[\text{c}],i}+d_{[\text{p}],i})R$  and the sum rate is given by 
\begin{align} \label{eq:sum-rate}
R_{\text{sum}}&= R_1+R_2 \nonumber \\
&=(d_{[\text{c}],1}+d_{[\text{p}],1}+d_{[\text{c}],2}+d_{[\text{p}],2})R.
\end{align}

For notional convenience, we define the set $d_{\text{stream}}$ as $d_{\text{stream}}=\{d_{[\text{c}],1},d_{[\text{p}],1},d_{[\text{c}],2},d_{[\text{p}],2}\}$ in the rest of the paper.

\subsubsection{Transmit beamforming} To  send multiple data streams simultaneously with multiple transmit antennas, we design the transmit signal vector of transmitter $i\in\{1,2\}$ at time slot $t\in[1:n]$ as 
\begin{align}\label{eq:tx signal}
\bar{\mathbf{x}}_i(t)=\mathbf{V}_{{[\text{c}],i}}\mathbf{s}_{[\text{c}],i}(t)+\mathbf{V}_{{[\text{p}],i}}\mathbf{s}_{[\text{p}],i}(t),
\end{align}
where  
\begin{align*}
\mathbf{s}_{[\text{c}],i}(t)&=\left[\begin{array}{ccccccc}s_{[\text{c}],i,1}(t) & s_{[\text{c}],i,2}(t) & \cdots & s_{[\text{c}],i,d_{[\text{c}],i}}(t) \end{array}\right]^\dagger\in \mathbb{R}^{d_{[\text{c}],i}\times 1}, \nonumber \\
\mathbf{s}_{[\text{p}],i}(t)&=\left[\begin{array}{ccccccc}s_{[\text{p}],i,1}(t) & s_{[\text{p}],i,2}(t) & \cdots & s_{[\text{p}],i,d_{[\text{p}],i}}(t) \end{array}\right]^\dagger \in \mathbb{R}^{d_{[\text{p}],i}\times 1}, 
\end{align*}
$s_{[\text{c}],i,j}(t)$ and $s_{[\text{p}],i,j}(t)$ are the $t$th elements in $\mathbf{s}_{[\text{c}],i,j}$ and $\mathbf{s}_{[\text{p}],i,j}$, respectively, and $\mathbf{V}_{{[\text{c}],i}}\in \mathbb{R}^{2M_\text{T}\times d_{[\text{c}],i}}$ and $\mathbf{V}_{{[\text{p}],i}}\in \mathbb{R}^{2M_\text{T}\times d_{[\text{p}],i}}$ are beamforming matrices that convey common and private streams of transmitter $i$, respectively, where all column vectors in $\mathbf{V}_{{[\text{c}],i}}$ and $\mathbf{V}_{{[\text{p}],i}}$ are linearly independent, i.e., 
\begin{align}\label{condition 1}
\text{rank}\left(\left[\begin{array}{ccccc}\mathbf{V}_{{[\text{c}],i}} & \mathbf{V}_{{[\text{p}],i}}\end{array}\right]\right)&=d_{[\text{c}],i}+d_{[\text{p}],i}\nonumber \\
&=d_i\nonumber \\
&\leq \min \{2M_{\text{T}}, 2M_{\text{R}}\}.
\end{align}
In addition, we choose $d_{\text{stream}}$ to satisfy
\begin{align}\label{condition 2}
d_{[\text{c}],1}+d_{[\text{c}],2}+d_{[\text{p}],i} \leq 2M_{\text{R}}
\end{align}
for all $i\in\{1,2\}$. Note that this condition is required to perform IF sum decoding for the intended streams and the other user's common streams at each receiver.

 Furthermore, the column vectors in $\mathbf{V}_{{[\text{c}],i}}$ and $\mathbf{V}_{{[\text{p}],i}}$ are properly normalized to satisfy the transmit power constraint $P$, i.e., their norms are given by one.  The detailed design of beamforming matrices  $\mathbf{V}_{{[\text{c}],i}}$ and $\mathbf{V}_{{[\text{p}],i}}$ suitable for each CSIT assumption will be discussed later in Section~\ref{Tx beamforming design}.
%and the decision rule for the set $(d_{[\text{c}],1},d_{[\text{p}],1},d_{[\text{c}],2},d_{[\text{p}],2})$

\subsection{Receiver Side}
\subsubsection{Linear filtering}
Since the transmit signal vector of  transmitter $i$ is set to~\eqref{eq:tx signal}, for given $\mathbf{V}_{{[\text{c}],i}}$ and $\mathbf{V}_{{[\text{p}],i}}$, the received signal vector of receiver $i$ at time slot $t\in[1:n]$ is given as 
\begin{align}
\bar{\mathbf{y}}_i(t)&=\bar{\mathbf{H}}_{i,i}\bar{\mathbf{x}}_i(t)+\bar{\mathbf{H}}_{i,j}\bar{\mathbf{x}}_j(t)+\bar{\mathbf{z}}_i(t)  \nonumber \\\displaybreak[1]
&=\bar{\mathbf{H}}_{i,i}\left(\mathbf{V}_{{[\text{c}],i}}\mathbf{s}_{[\text{c}],i}(t)+\mathbf{V}_{{[\text{p}],i}}\mathbf{s}_{[\text{p}],i}(t)\right)+\bar{\mathbf{H}}_{i,j}\left(\mathbf{V}_{{[\text{c}],j}}\mathbf{s}_{[\text{c}],j}(t)+\mathbf{V}_{{[\text{p}],j}}\mathbf{s}_{[\text{p}],j}(t)\right)+\bar{\mathbf{z}}_i(t)  \nonumber \\ \displaybreak[1]
&=\tilde{\mathbf{H}}_{[\text{d}],i} \left[\begin{array}{ccccc}\mathbf{s}_{[\text{c}],i}(t) \\ \mathbf{s}_{[\text{p}],i}(t)\\ \mathbf{s}_{[\text{c}],j}(t) \end{array}\right]+\tilde{\mathbf{H}}_{[\text{i}],i}\mathbf{s}_{[\text{p}],j}(t)+\bar{\mathbf{z}}_i(t)\nonumber \\
&=\tilde{\mathbf{H}}_{[\text{d}],i}\mathbf{s}_{[\text{d}],i}(t)+\tilde{\mathbf{H}}_{[\text{i}],i}\mathbf{s}_{[\text{p}],j}(t)+\bar{\mathbf{z}}_i(t),
\end{align}
where $i,j \in\{1,2\}$ and $i\neq j$. Here
\begin{align*}
\tilde{\mathbf{H}}_{[\text{d}],i} &=\left[\begin{array}{ccccccc}\bar{\mathbf{H}}_{i,i}\mathbf{V}_{{[\text{c}],i}}& \bar{\mathbf{H}}_{i,i}\mathbf{V}_{{[\text{p}],i}} & \bar{\mathbf{H}}_{i,j}\mathbf{V}_{{[\text{c}],j}} \end{array}\right]\in \mathbb{R}^{2M_{\text{R}}\times \left(d_{[\text{c}],i}+d_{[\text{p}],i}+d_{[\text{c}],j}\right)}
\end{align*}
represents the effective desired channel matrix observed at receiver $i$, 
\begin{align*}
\tilde{\mathbf{H}}_{[\text{i}],i}&=\bar{\mathbf{H}}_{i,j}\mathbf{V}_{{[\text{p}],j}}\in \mathbb{R}^{2M_{\text{R}}\times d_{[\text{p}],j}}
\end{align*} 
represents the effective interfering channel matrix observed at receiver $i$, and 
\begin{align*}
\mathbf{s}_{[\text{d}],i}(t)=[\mathbf{s}_{[\text{c}],i}^\dagger (t)~\mathbf{s}_{[\text{p}],i}^\dagger (t)~ \mathbf{s}_{[\text{c}],j}^\dagger (t)]^\dagger \in\mathbb{R}^{\left(d_{[\text{c}],i}+d_{[\text{p}],i}+d_{[\text{c}],j}\right)\times 1}.
\end{align*} 
In addition, for notational simplicity, let $d_{[\text{d}] ,i}=d_{[\text{c}],i}+d_{[\text{p}],i}+d_{[\text{c}],j}$, which 
is the total number of streams that receiver $i$ attempts to recover. 
Then, based on $\bar{\mathbf{Y}}_i=[\begin{array}{ccccc}\bar{\mathbf{y}}_i(1) & \bar{\mathbf{y}}_i(2) &\cdots& \bar{\mathbf{y}}_i(n) \end{array}]\in \mathbb{R}^{2M_{\text{R}}\times n}$, receiver $i$ tries to recover the data streams associated with 
$$\mathbf{S}_{[\text{d}],i}=[\begin{array}{ccccc}\mathbf{s}_{\text{[d]},i}(1) & \mathbf{s}_{\text{[d]},i}(2) &\cdots& \mathbf{s}_{\text{[d]},i}(n) \end{array}]\in \mathbb{R}^{d_{[\text{d}] ,i}\times n}$$
by applying IF sum decoding while treating
$$\mathbf{S}_{[\text{p}],j}=[\begin{array}{ccccc}\mathbf{s}_{\text{[p]},j}(1) & \mathbf{s}_{\text{[p]},j}(2) &\cdots& \mathbf{s}_{\text{[p]},j}(n) \end{array}]\in \mathbb{R}^{d_{[\text{p}] ,j}\times n}$$
as noise. To this end, receiver $i$ applies a linear filter $\mathbf{F}_{\text{IF},i}\in\mathbb{R}^{d_{[\text{d}] ,i}\times 2M_{\text{R}}}$ to the received vector $\bar{\mathbf{y}}_i(t)$ for each time slot $t\in[1:n]$ to obtain
\begin{align}\label{eq:IF-filter}
\tilde{\mathbf{y}}_i(t)&=\mathbf{F}_{\text{IF},i}\bar{\mathbf{y}}_i(t)\nonumber\\
&=\mathbf{F}_{\text{IF},i}\left(\tilde{\mathbf{H}}_{[\text{d}],i}\mathbf{s}_{[\text{d}],i}(t)+\tilde{\mathbf{H}}_{[\text{i}],i}\mathbf{s}_{[\text{p}],j}(t)\right)+\mathbf{F}_{\text{IF},i}\bar{\mathbf{z}}_i(t) \displaybreak[1]\nonumber\\
&=\mathbf{A}_i \mathbf{s}_{[\text{d}],i}(t)+\underbrace{\left(\mathbf{F}_{\text{IF},i}\tilde{\mathbf{H}}_{[\text{d}],i}-\mathbf{A}_i\right)\mathbf{s}_{[\text{d}],i}(t)+\mathbf{F}_{\text{IF},i}\tilde{\mathbf{H}}_{[\text{i}],i}\mathbf{s}_{[\text{p}],j}(t)+ \mathbf{F}_{\text{IF},i}\bar{\mathbf{z}}_i(t)}_{\textrm{Effective noise}}\displaybreak[1]\nonumber \\
&=\mathbf{A}_i \mathbf{s}_{[\text{d}],i}(t)+\tilde{\mathbf{z}}_i(t),
\end{align}
where
\begin{align}\label{IF_filter}
&\mathbf{F}_{\text{IF},i}=P\mathbf{A}_i{\tilde{\mathbf{H}}_{[\text{d}],i}}^\dagger{\left(\mathbf{I}_{2 M_{\text{R}}}+P\tilde{\mathbf{H}}_{[\text{d}],i}{\tilde{\mathbf{H}}_{[\text{d}],i}}^{\dagger}
+P\tilde{\mathbf{H}}_{[\text{i}],i}{\tilde{\mathbf{H}}_{[\text{i}],i}}^{\dagger}\right)}^{-1},   \\
&\tilde{\mathbf{z}}_i(t)=\left(\left(\mathbf{F}_{\text{IF},i}\tilde{\mathbf{H}}_{[\text{d}],i}-\mathbf{A}_i\right)\mathbf{s}_{[\text{d}],i}(t)+\mathbf{F}_{\text{IF},i}\tilde{\mathbf{H}}_{[\text{i}],i}\mathbf{s}_{[\text{p}],j}(t)+ \mathbf{F}_{\text{IF},i}\bar{\mathbf{z}}_i(t)\right),
\end{align}
and $\mathbf{A}_i\in \mathbb{Z}^{d_{[\text{d}] ,i}\times d_{[\text{d}] ,i}}$ is a full-rank integer matrix chosen such that the effective noise in~\eqref{eq:IF-filter} is minimized. By extending the results in~\cite[Theorem 4 and Section VII-B]{Zhan:14}, it can be seen that near-optimal $\mathbf{A}_i$ can be found by solving the following optimization problem\cite{Zhan:14,Bakoury:15} with approximate search algorithms such as  Lenstra--Lenstra--Lovasz (LLL) algorithm~\cite{LLL},
\begin{align}\label{eq:Integer_matrix_search_channel_variation}
\underset{\mathbf{A}_i\in  \mathbb{Z}^{d_{\text{[d]},i}\times d_{\text{[d]},i}},~
\textrm{rank}(\mathbf{A}_i)=d_{\text{[d]},i}}{\arg\min}{\max_{m\in[1:d_{\text{[d]},i}]}}\|\mathbf{{\Gamma}}_i^{-1/2}\mathbf{{U}}_i^\dagger \mathbf{a}_{i,m}\|^2,
\end{align}
%where $\mathbf{{V}}$ and $\mathbf{{D}}$ are obtained from the eigendecomposition of  $\mathbf{{Q}}=\frac{1}{n}\sum_{t=1}^n(\mathbf{I}_{2M}+P {\bar{\mathbf{H}}}^\dagger(t){\bar{\mathbf{H}}(t)})$, i.e., $\mathbf{Q}=\mathbf{V}\mathbf{D}\mathbf{V}^\dagger$,
where $\mathbf{U}_i$ is an orthogonal matrix consisted of the eigenvectors of
\begin{align}\label{Q}
\mathbf{Q}_i=\left(\mathbf{I}_{2 M_{\text{R}}}-P \mathbf{H}_{[\text{d}],i}^\dagger
{\left(\mathbf{I}_{2 M_{\text{R}}}+P\tilde{\mathbf{H}}_{[\text{d}],i}{\tilde{\mathbf{H}}_{[\text{d}],i}}^{\dagger}
+P\tilde{\mathbf{H}}_{[\text{i}],i}{\tilde{\mathbf{H}}_{[\text{i}],i}}^{\dagger}
\right)}^{-1} \mathbf{H}_{[\text{d}],i}\right)^{-1}
\end{align}
as its columns, $\mathbf{\Gamma}_i$ is a diagonal matrix whose entries are the eigenvalues of $\mathbf{Q}_i$, and $\mathbf{a}^\dagger_{i,m}\in \mathbb{R}^{1\times d_{[\text{d}] ,i}}$ is the $m$th row vector in $\mathbf{A}_i$.
Moreover, after obtaining $\mathbf{A}_i$, the optimal linear filter $\mathbf{F}_{\text{IF},i}$
can be calculated in a closed form as~\eqref{IF_filter}.

\subsubsection{IF sum decoding and achievable rate}
After obtaining $\tilde{\mathbf{Y}}_i=[\begin{array}{ccccc}\tilde{\mathbf{y}}_i(1) & \tilde{\mathbf{y}}_i(2) &\cdots& \tilde{\mathbf{y}}_i(n) \end{array}]\in \mathbb{R}^{d_{[\text{d}] ,i}\times n}$ over the block of $n$ transmissions,  receiver $i$ first adds the dither matrix $\mathbf{D}_i$ to $\tilde{\mathbf{Y}}_i$ and performs the modulo-$\Lambda_c$ operation to get~\cite{Ordentlich:conf_13}
\begin{align} \label{eq:in_out}
\tilde{\tilde{\mathbf{Y}}}_i&=\left(\tilde{\mathbf{Y}}_i+ \mathbf{A}_i\mathbf{D}_i\right)\mod \Lambda_c \nonumber \\
&=\left(\mathbf{A}_i \mathbf{S}_{[\text{d}],i}+ \mathbf{A}_i\mathbf{D}_i+\tilde{\mathbf{Z}}_i\right)\mod \Lambda_c \nonumber \\
&=\left(\mathbf{A}_i \mathbf{B}_{[\text{d}],i}+\tilde{\mathbf{Z}}_i\right)\mod \Lambda_c \nonumber \\
&=\left(\mathbf{C}_i+\tilde{\mathbf{Z}}_i\right)\mod \Lambda_c,
\end{align}
where 
\begin{align*}
&\tilde{\mathbf{Y}}_i=\left[\begin{array}{ccccccccccccc}\tilde{\mathbf{y}}_i(1)& \tilde{\mathbf{y}}_i(2)&\cdots&\tilde{\mathbf{y}}_i(n)\end{array}\right] \in \mathbb{R}^{d_{[\text{d}] ,i}\times n},\\
&\mathbf{D}_i=\left[\begin{array}{ccccccccccccc} \mathbf{d}_{[\text{c}],i,1}^\dagger & \cdots& \mathbf{d}_{[\text{c}],i,d_{\text{[c]},i}}^\dagger  &  \mathbf{d}_{[\text{p}],i,1}^\dagger & \cdots& \mathbf{d}_{[\text{p}],i,d_{\text{[p]},i}}^\dagger& \mathbf{d}_{[\text{c}],j,1}^\dagger & \cdots& \mathbf{d}_{[\text{c}],j,d_{\text{[c]},j}}^\dagger  \end{array}\right]^\dagger \in \mathbb{R}^{d_{[\text{d}] ,i}\times n},\\
&\tilde{\mathbf{Z}}_i=[\begin{array}{ccccc}\tilde{\mathbf{z}}_i(1) & \tilde{\mathbf{z}}_i(2) &\cdots& \tilde{\mathbf{z}}_i(n) \end{array}]\in \mathbb{R}^{d_{[\text{d}] ,i}\times n},\\
&\mathbf{B}_{[\text{d}],i}=\left[\begin{array}{ccccccccccccc} \mathbf{b}_{[\text{c}],i,1}^\dagger & \cdots& \mathbf{b}_{[\text{c}],i,d_{\text{[c]},i}}^\dagger  &  \mathbf{b}_{[\text{p}],i,1}^\dagger & \cdots& \mathbf{b}_{[\text{p}],i,d_{\text{[p]},i}}^\dagger& \mathbf{b}_{[\text{c}],j,1}^\dagger & \cdots& \mathbf{b}_{[\text{c}],j,d_{\text{[c]},j}}^\dagger  \end{array}\right]^\dagger \in \mathbb{R}^{d_{[\text{d}] ,i}\times n},
\end{align*}
and $\mathbf{C}_i=\left(\mathbf{A}_i \mathbf{B}_{[\text{d}],i}\right) \mod \Lambda_c$ for $i\neq j$. Observe that $\mathbf{B}_{[\text{d}],i}=\mathbf{S}_{[\text{d}],i}+\mathbf{D}_i$ and each row of $\mathbf{B}_{[\text{d}],i}$ is a codeword from the same lattice codebook $\mathcal{C}$, as described in Section~\ref{encoding}. Since any integer-linear combination of lattice codewords over $\Lambda_c$ from the same codebook $\mathcal{C}$ is also a codeword included in $\mathcal{C}$ due to the linearity property of lattice codes, each row of $\mathbf{C}_i$ is itself a codeword in $\mathcal{C}$, which can be decoded within noise tolerance. In addition, the additive noise $\tilde{\mathbf{Z}}_i$ is independent of $\mathbf{C}_i$ due to the fact that $\mathbf{B}_{[\text{d}],i}$ is independent of $\mathbf{S}_{[\text{d}],i}$, $\mathbf{S}_{[\text{p}],j}$, and $\bar{\mathbf{Z}}_i$, where 
\begin{align*}
\bar{\mathbf{Z}}_i=[\begin{array}{ccccc}\bar{\mathbf{z}}_i(1) & \bar{\mathbf{z}}_i(2) &\cdots& \bar{\mathbf{z}}_i(n) \end{array}]\in \mathbb{R}^{2M_{\text{R}} \times n}
\end{align*}
From \eqref{eq:in_out}, let  $\tilde{\tilde{\mathbf{y}}}_{i,m}$, $\mathbf{c}_{i,m}$, and $\tilde{\mathbf{z}}_{i,m}$ denote the $m$th row vector in $\tilde{\tilde{\mathbf{Y}}}_{i}$, $\mathbf{C}_i$, and $\tilde{\mathbf{Z}}_i$, respectively, where $m\in [1:d_{\text{[d]},i}]$. Then we have
\begin{align}
\tilde{\tilde{\mathbf{y}}}_{i,m}=\left(\mathbf{c}_{i,m}+\tilde{\mathbf{z}}_{i,m}\right) \mod \Lambda_c.
\end{align}
Based on $\tilde{\tilde{\mathbf{y}}}_{i,m}$, receiver $i$ attempts to separately decode $\mathbf{c}_{i,m}$  for each $m\in [1:d_{\text{[d]},i}]$ in the presence of the effective noise vector  $\tilde{\mathbf{z}}_{i,m}$~\cite{Zhan:14,Ordentlich:conf_13,Ordentlich:14,Ordentlich:15,Nazer:11}. Let $\hat{\mathbf{c}}_{i,m}$ denote the decoded version of  $\mathbf{c}_{i,m}$ and
\begin{align*}
\hat{\mathbf{C}}_i=\left[\begin{array}{ccccccccccccc}\hat{\mathbf{c}}_{i,1}^\dagger & \hat{\mathbf{c}}_{i,2}^\dagger& \cdots& \hat{\mathbf{c}}_{i,d_{\text{[d]},i}}^\dagger  \end{array}\right]^\dagger \in \mathbb{R}^{d_{[\text{d}] ,i}\times n}.
\end{align*}
Then the original streams in $\mathbf{B}_{[\text{d}],i}$ can be recovered from the decoded output $\hat{\mathbf{C}}_i$ as
\begin{align}
\hat{\mathbf{B}}_{[\text{d}],i}=\left(\mathbf{A}_{\Lambda_c}^{-1} \hat{\mathbf{C}}_i\right) \mod \Lambda_c, 
\end{align}
where  $\mathbf{A}_{\Lambda_c}=(\mathbf{A}) \mod \Lambda_c$ and the matrix inversion is taken over $\Lambda_c$.
Recall that receiver $i$ tries to recover all the streams in $\mathbf{B}_{[\text{d}],i}$, i.e., it recovers not only its intended streams sent by transmitter $i$,  but also the common streams sent by transmitter $j$ under the proposed scheme, where $j\neq i$. As in~\cite{Zhan:14,Ordentlich:conf_13,Ordentlich:14,Ordentlich:15,Nazer:11}, the decoding of $\mathbf{c}_{i,m}$ is successful with probability approaching one as $n\rightarrow \infty$ if 
\begin{align}
R<\frac{1}{2}\log^{+}\left(\frac{P}{\sigma^2_{i,m}}\right),
\end{align}
where $\sigma^2_{i,m}$ denotes the variance of $\tilde{\mathbf{z}}_{i,m}$ given by $\sigma^2_{i,m}=\frac{1}{n}\mathbb{E}\|\tilde{\mathbf{z}}_{i,m}\|^2$. Since all integer-linear combinations  $\{\mathbf{c}_{i,m}\}_{m\in [1:d_{\text{[d]},i}]}$  should be successfully decoded in order to recover $\mathbf{B}_{[\text{d}],i}$ for all $i\in \{1,2\}$, the achievable rate for each data stream is given by 
\begin{align}
R<\min_{i\in\{1,2\},m\in[1:d_{\text{[d]},i}]}\frac{1}{2}\log^+\left(\frac{P}{\sigma^2_{i,m}}\right).
\end{align}
As a result, from~\eqref{eq:sum-rate}, the overall achievable rates for receivers 1 and 2 and the corresponding sum rate are given by
\begin{align}\label{achievable rate}
R_1&<\left(d_{[\text{c}],1}+d_{[\text{p}],1}\right)\left(\min_{i\in\{1,2\},m\in[1:d_{\text{[d]},i}]}\frac{1}{2}\log^+\left(\frac{P}{\sigma^2_{i,m}}\right)\right),\nonumber \\
R_2&<\left(d_{[\text{c}],2}+d_{[\text{p}],2}\right)\left(\min_{i\in\{1,2\},m\in[1:d_{\text{[d]},i}]}\frac{1}{2}\log^+\left(\frac{P}{\sigma^2_{i,m}}\right)\right),\nonumber \\
R_1+R_2&<\left(d_{[\text{c}],1}+d_{[\text{p}],1}+d_{[\text{c}],2}+d_{[\text{p}],2}\right)\left(\min_{i\in\{1,2\},m\in[1:d_{\text{[d]},i}]}\frac{1}{2}\log^+\left(\frac{P}{\sigma^2_{i,m}}\right)\right).
\end{align}

Note that we can rewrite the achievable rates~\eqref{achievable rate} in a different form. Let us consider the following Cholesky decomposition:
\begin{align}\label{eq:Cholesky_1}
\mathbf{A}_i\mathbf{Q}_i^{-1}\mathbf{A}_i^\dagger=\mathbf{L}_i\mathbf{L}_i^\dagger,
\end{align}
where $\mathbf{Q}_i$ is given as~\eqref{Q} and $\mathbf{L}_i\in\mathbb{R}^{d_{\text{[d]},i}\times d_{\text{[d]},i}}$ is a lower triangular matrix. Then it can be shown that the effective noise variance $\sigma^2_{i,m}$ is given by
\begin{align}\label{effective noise_IF}
\sigma^2_{i,m}=P\sum_{k=1}^m l^2_{i,k,m},
\end{align}
where $ l_{i,k,m}$ is the $(k,m)$th element of $\mathbf{L}_i$. To avoid duplication of explanation, we refer to~\cite[Section III]{Ordentlich:conf_13} for detailed derivation. Finally, the achievable rates~\eqref{achievable rate} are rewritten as 
\begin{align}\label{achievable rate2}
R_1&<\left(d_{[\text{c}],1}+d_{[\text{p}],1}\right)\left(\min_{i\in\{1,2\},m\in[1:d_{\text{[d]},i}]}\frac{1}{2}\log^+\left(\frac{1}{{\sum_{k=1}^m l^2_{i,k,m}}}\right)\right),\nonumber \\
R_2&<\left(d_{[\text{c}],2}+d_{[\text{p}],2}\right)\left(\min_{i\in\{1,2\},m\in[1:d_{\text{[d]},i}]}\frac{1}{2}\log^+\left(\frac{1}{\sum_{k=1}^m l^2_{i,k,m}}\right)\right),\nonumber \\
R_1+R_2&<\left(d_{[\text{c}],1}+d_{[\text{p}],1}+d_{[\text{c}],2}+d_{[\text{p}],2}\right)\left(\min_{i\in\{1,2\},m\in[1:d_{\text{[d]},i}]}\frac{1}{2}\log^+\left(\frac{1}{\sum_{k=1}^m l^2_{i,k,m}}\right)\right).
\end{align}

\begin{remark}
Recall that $\mathbf{A}_1$ and $\mathbf{A}_2$ can be constructed by solving the optimization problem \eqref{eq:Integer_matrix_search_channel_variation}. Then, from \eqref{eq:Cholesky_1},  $\mathbf{L}_i$ can be obtained by using  such $\mathbf{A}_i$ and $\mathbf{Q}_i$ given in \eqref{Q} and, as a result, the right-hand side terms in \eqref{achievable rate2}, i.e.,
\begin{align*}
\min_{i\in\{1,2\},m\in[1:d_{\text{[d]},i}]}\frac{1}{2}\log^+\left(\frac{1}{{\sum_{k=1}^m l^2_{i,k,m}}}\right),\nonumber \\
\min_{i\in\{1,2\},m\in[1:d_{\text{[d]},i}]}\frac{1}{2}\log^+\left(\frac{1}{\sum_{k=1}^m l^2_{i,k,m}}\right),\nonumber \\
\min_{i\in\{1,2\},m\in[1:d_{\text{[d]},i}]}\frac{1}{2}\log^+\left(\frac{1}{\sum_{k=1}^m l^2_{i,k,m}}\right)
\end{align*}
are determined. 
Therefore, \textcolor{black}{the rest of the optimization process is to find an appropriate set of streams $d_{\text{stream}}$ and design beamforming vectors $\{\mathbf{V}_{{[\text{c}],1}}, \mathbf{V}_{{[\text{p}],1}},\mathbf{V}_{{[\text{c}],2}}, \mathbf{V}_{{[\text{p}],2}}\}$ for a given CSIT assumption.} \hfill$\lozenge$
\end{remark}

\subsubsection{IF sum decoding with SIC}
So far, we focused on IF sum decoding without SIC. By combining IF sum decoding with SIC, namely, \emph{successive IF}~\cite{Ordentlich:conf_13}, the achievable rates~\eqref{achievable rate} or~\eqref{achievable rate2} can be further improved. In contrast to the basic IF receiver that recovers integer-linear combinations of codewords in parallel  as described earlier,  the successive IF receiver attempts to decode them one-by-one sequentially.
To be specific, after linear combination $\mathbf{c}_{i,m}$ is recovered, the receiver estimates the  effective noise corresponding to that summation, i.e., it estimates $\tilde{\mathbf{z}}_{i,m}$ based on $\tilde{\tilde{\mathbf{y}}}_{i,m}$ and the decoding output $\hat{\mathbf{c}}_{i,m}$. The receiver then performs SIC of the estimated noise for all $\{\tilde{\tilde{\mathbf{y}}}_{i,j}\}_{j>m}$ to reduce the effective noise for the remaining integer-linear combinations that have not yet been decoded. 
To state the achievable rate of each user that can be obtained through the successive IF scheme, let $\mathbf{L}_{\text{SIC},i}\in\mathbb{R}^{d_{\text{[d]},i}\times d_{\text{[d]},i}}$ denote the lower triangular matrix obtained from the following Cholesky decomposition:
\begin{align}\label{Cholesky_2}
\mathbf{A}_{\text{SIC},i}\mathbf{Q}_i^{-1}\mathbf{A}_{\text{SIC},i}^\dagger=\mathbf{L}_{\text{SIC},i}\mathbf{L}_{\text{SIC},i}^\dagger,
\end{align}
where $\mathbf{A}_{\text{SIC},i}\in\mathbb{Z}^{d_{[\text{d}] ,i}\times d_{[\text{d}] ,i}}$ is a full-rank integer matrix and $\mathbf{Q}_i$ is given as~\eqref{Q}. 
Here, a proper integer matrix of the successive IF scheme for receiver $i$,  denoted by $\mathbf{A}_{\text{SIC},i}$, can be obtained by solving the following optimization problem~\cite{Ordentlich:conf_13}
\begin{align}\label{eq:Integer_matrix_search_channel_variation_2}
\underset{\mathbf{A}_{\text{SIC},i}\in  \mathbb{Z}^{d_{\text{[d]},i}\times d_{\text{[d]},i}},~
\textrm{rank}(\mathbf{A}_{\text{SIC},i})=d_{\text{[d]},i}}{\arg\min}{\max_{ m\in[1:d_{\text{[d]},i}]}}l^2_{\text{SIC},i,m,m}
\end{align}
instead of solving~\eqref{eq:Integer_matrix_search_channel_variation}, where  $l_{\text{SIC},i.k,m}$ denotes the $(k,m)$th element of $\mathbf{L}_{\text{SIC},i}$.
Then the achievable rates of the successive IF scheme are given by
\begin{align}\label{achievable rate3}
R_1&<\left(d_{[\text{c}],1}+d_{[\text{p}],1}\right)\left(\min_{i\in\{1,2\},m\in[1:d_{\text{[d]},i}]}\frac{1}{2}\log^+\left(1 \middle/ l^2_{\text{SIC},i,m,m}\right)\right),\nonumber \\
R_2&<\left(d_{[\text{c}],2}+d_{[\text{p}],2}\right)\left(\min_{i\in\{1,2\},m\in[1:d_{\text{[d]},i}]}\frac{1}{2}\log^+\left(1 \middle/ l^2_{\text{SIC},i,m,m}\right)\right),\nonumber \\
R_1+R_2&<\left(d_{[\text{c}],1}+d_{[\text{p}],1}+d_{[\text{c}],2}+d_{[\text{p}],2}\right)\left(\min_{i\in\{1,2\},m\in[1:d_{\text{[d]},i}]}\frac{1}{2}\log^+\left(1 \middle/ l^2_{\text{SIC},{i,m,m}}\right)\right).
\end{align}
That is, the effective noise variance $\sigma^2_{i,m}$ of the successive IF scheme becomes 
\begin{align}\label{effective noise_successive IF}
\sigma^2_{i,m}= P l^2_{\text{SIC}, i,m,m}, 
\end{align}
which is clearly smaller than that in~\eqref{effective noise_IF} even when $\mathbf{A}_{\text{SIC},i}=\mathbf{A}_{i}$. As a result, the successive IF receiver can outperform the basic IF receiver due to this noise reduction. For more detailed derivations, see~\cite{Ordentlich:conf_13}. 

\subsection{Design of Transmit Beamforming}\label{Tx beamforming design}
The achievable rates~\eqref{achievable rate2} or \eqref{achievable rate3} depend on how to design transmit beamforming matrices  $\mathbf{V}_{{[\text{c}],i}}$ and $\mathbf{V}_{{[\text{p}],i}}$, $i\in\{1,2\}$. However, finding the optimal beamforming matrices that maximize the achievable rates is very difficult to solve because the considered optimization problem belongs to non-convex integer programming.\footnote{We also note that the capacity of the two-user MIMO interference channel is not yet known in general.} Therefore, in this paper, we focus on developing a simple universal beamforming scheme that can provide good performance over a  wide range of channel matrices with low complexity suitable for each CSIT assumption.
\subsubsection{No CSIT case} When  CSI is not available at the transmitter side, we set the transmit beamforming matrices as
\begin{align}\label{no CSIT}
&\mathbf{V}_{{[\text{c}],i}}=[\mathbf{I}_{2M_\text{T}}]_{1:d_{\text{[c]},i}},\nonumber\\
&\mathbf{V}_{{[\text{p}],i}}=[\mathbf{I}_{2M_\text{T}}]_{d_{\text{[c]}}+1:d_{\text{[c]}}+d_{\text{[p]},i}}
\end{align}
for all $i\in\{1,2\}$ by utilizing the identity matrix. Then, based on the above transmit beamforming matrices~\eqref{no CSIT}, the set $d_{\text{stream}}=\{d_{[\text{c}],1},d_{[\text{p}],1},d_{[\text{c}],2},d_{[\text{p}],2}\}$ is numerically optimized to maximize the achievable rates~\eqref{achievable rate2} or \eqref{achievable rate3}  while satisfying the conditions~\eqref{condition 1} and~\eqref{condition 2}.\footnote{We assume that the optimal set $\{d_{[\text{c}],1},d_{[\text{p}],1},d_{[\text{c}],2},d_{[\text{p}],2}\}$ is searched by each receiver and reported to the corresponding transmitter before communication for all CSIT assumptions. }
\subsubsection{{Partial CSIT case}}\label{Sec:Partial CSIT Case} Now consider the case in which only the CSI of its own desired link is available at each transmitter, i.e., 
only the coefficients in $\bar{\mathbf{H}}_{i,i}$ is known to transmitter $i\in\{1,2\}$. Consider the singular value decomposition (SVD) of $\bar{\mathbf{H}}_{i,i}$ given by 
\begin{align}\label{SVD}
\bar{\mathbf{H}}_{i,i}=\mathbf{G}_{i,i}\mathbf{\Sigma}_{i,i}\mathbf{E}^\dagger_{i,i}
\end{align}
where $\mathbf{G}_{i,i}\in \mathbb{R}^{2M_{\text R}\times 2M_{\text R}}$ and $\mathbf{E}_{i,i}\in \mathbb{R}^{2M_{\text T}\times 2M_{\text T}}$ are unitary matrices and $\mathbf{\Sigma}_{i,i}\in \mathbb{R}^{2M_{\text R}\times 2M_{\text T}}$ is a rectangular diagonal matrix with non-negative real diagonal values. Then, to utilize the CSI of $\bar{\mathbf{H}}_{i,i}$, we design the transmit beamforming matrices by using the SVD of $\bar{\mathbf{H}}_{i,i}$ as
\begin{align}\label{partial CSIT}
&\mathbf{V}_{{[\text{c}],i}}=[\mathbf{E}_{i,i}]_{1:d_{\text{[c]},i}},\nonumber\\
&\mathbf{V}_{{[\text{p}],i}}=[\mathbf{E}_{i,i}]_{d_{\text{[c]},i}+1:d_{\text{[c]},i}+d_{\text{[p]},i}}.
\end{align}
Note that SVD-based beamforming was also considered for a point-to-point MIMO channel with an IF receiver when CSI is known to both the transmitter and receiver,  i.e., the closed-loop communication scenario~\cite{Sakzad:15}. Specifically, it was shown that unitary precoded IF in conjunction with SVD can achieve full-diversity.\footnote{However, the optimal precoding for an IF receiver that maximizes the achievable rate has not yet been known even for the point-to-point channel.}

Then, based on the transmit beamforming matrices~\eqref{partial CSIT}, the optimal set of $d_{\text{stream}}$ that maximizes the achievable rates~\eqref{achievable rate2} or \eqref{achievable rate3} is numerically searched under the conditions~\eqref{condition 1} and~\eqref{condition 2}.

\subsubsection{Full CSIT case} \label{Sec:Full CSIT Case} Finally, consider the case in which full CSI is available at each transmitter, i.e., the coefficients in $\bar{\mathbf{H}}_{1,i}$ and $\bar{\mathbf{H}}_{2,i}$ are known to transmitter $i\in\{1,2\}$. Let us consider the SVDs of  $\bar{\mathbf{H}}_{i,i}$ and  $\bar{\mathbf{H}}_{j,i}$ given by $
\bar{\mathbf{H}}_{i,i}=\mathbf{G}_{i,i}\mathbf{\Sigma}_{i,i}\mathbf{E}^\dagger_{i,i}$ and 
$\bar{\mathbf{H}}_{j,i}=\mathbf{G}_{j,i}\mathbf{\Sigma}_{j,i}\mathbf{E}^\dagger_{j,i}$ in the same manner as in~\eqref{SVD}, where $j\neq i$. 
Additionally, we also consider the SVD of the concatenated matrix $[\begin{array}{ccc}\bar{\mathbf{H}}^\dagger_{i,i} & \bar{\mathbf{H}}^\dagger_{j,i} \end{array}]^\dagger$ given by 
$[\begin{array}{ccc}\bar{\mathbf{H}}^\dagger_{i,i} & \bar{\mathbf{H}}^\dagger_{j,i} \end{array}]^\dagger=\mathbf{G}_{i}\mathbf{\Sigma}_{i}\mathbf{E}^\dagger_{i}$. Then we design the transmit beamforming matrices as 
\begin{align}\label{full CSIT-common}
\mathbf{V}_{[\text{c}],i}=[\mathbf{E}_i]_{1:d_{\text{[c]},i}},
\end{align}
\begin{align}\label{full CSIT-private}
\mathbf{V}_{{[\text{p}],i}}=\begin{cases} [\mathbf{E}_{i,i}]_{1:d_{\text{[p]},i}} \quad \quad\quad \quad \quad \quad\quad \quad \quad \quad \mbox{if } M_\text{T} \leq M_\text{R}, \\
\left[\begin{array}{ccc}\gamma_i[\mathbf{E}_{i,i}]_{1:2M_\text{T}-2M_\text{R}}+(1-\gamma_i)[\mathbf{J}_{j,i}]_{1:2M_\text{T}-2M_\text{R}}& [\mathbf{E}_{i,i}]_{2M_\text{T}-2M_\text{R}+1: d_{\text{[p]},i}} \end{array} \right]\\
\quad \quad\quad \quad \quad \quad\quad \quad \quad \quad \quad\quad \quad ~~ \mbox{if } 0<2M_\text{T} - 2M_\text{R}\leq d_{\text{[p]},i}, \\
\gamma_i[\mathbf{E}_{i,i}]_{1:d_{\text{[p]},i}}+(1-\gamma_i)[\mathbf{J}_{j,i}]_{1:d_{\text{[p]},i}} \quad \mbox{otherwise,}  \end{cases}
\end{align}
where $\mathbf{J}_{j,i}\in \mathbb{R}^{2M_{\text{T}}\times \left(2M_{\text{T}}-2M_{\text{R}}\right)}$ denotes the null matrix of $\bar{\mathbf{H}}_{j,i}$ that satisfies $\bar{\mathbf{H}}_{j,i}\mathbf{J}_{j,i}=\mathbf{0}_{2M_{\text{R}}\times\left(2M_{\text{T}}-2M_{\text{R}}\right)}$ and $\text{rank}(\bar{\mathbf{H}}_{i,i}\mathbf{J}_{j,i})=\min\{2M_\text{R},2M_{\text{T}}-2M_{\text{R}} \}$ when $M_{\text{T}}>M_{\text{R}}$ and $0\leq \gamma_i\leq 1$  is a real number, which is numerically optimized assuming that the other parameters are given. Recall that the column vectors in $\mathbf{V}_{{[\text{c}],i}}$ and $\mathbf{V}_{{[\text{p}],i}}$ are properly normalized to satisfy the transmit power constraint $P$ for all $i=1,2$. 

 The motivation of the proposed beamforming construction~\eqref{full CSIT-common} is from the fact that common streams are required to be recovered by both receivers. Hence, the right unitary matrix of the concatenated matrix consisting of the desired and interfering links are employed to convey common streams. On the other hand, since private streams are only    recovered by the dedicated receiver while interfering to the other receiver, ZF beamforming is used to null out interference to the unintended receiver in addition to the SVD beamforming in~\eqref{full CSIT-private}. More specifically, a hybrid scheme, which is a simple linear combination of the ZF and SVD beamforming schemes, is proposed in~\eqref{full CSIT-private} as a compromise between the two approaches.

Then, based on the transmit beamforming matrices~\eqref{full CSIT-common} and~\eqref{full CSIT-private}, the set $d_{\text{stream}}$ and $(\gamma_1,\gamma_2)$ are optimized to maximize the achievable rates~\eqref{achievable rate2} or \eqref{achievable rate3} while satisfying the conditions~\eqref{condition 1} and~\eqref{condition 2}.

 \subsection{Remarks} \label{subsec:remarks}
 \subsubsection{Comparison with previous works and implementation via practical codes in practical environments}
 By restricting $\mathbf{A}_{\text{SIC},i}$  in~\eqref{Cholesky_2} to an identity matrix for all $i\in\{1,2\}$, the proposed successive IF scheme becomes the conventional scheme with MMSE-SIC receivers, in which the operation at the transmitter side is the same as that of the proposed successive IF while MMSE-SIC operation is performed instead of successive IF sum decoding at the receiver side. Similarly, the proposed IF scheme recovers the conventional scheme with MMSE receivers by  restricting $\mathbf{A}_{i}$  in~\eqref{eq:Cholesky_1} to an identity matrix for all $i\in\{1,2\}$. Moreover, if we set $d_{[\text{p}],i}=d_i=\min \{2M_{\text{T}}, 2M_{\text{R}}\}$ and $d_{[\text{c}],i}=0$ for all $i\in\{1,2\}$, i.e., sending private streams only without any rank adaptation, then the proposed IF scheme recovers the previous interference management scheme with IF receivers proposed in~\cite[Section VII-C]{Zhan:14}. 
Therefore, the proposed IF framework provides a general coding strategy including  the conventional MMSE-SIC receivers, MMSE receivers, and IF receivers as special cases.   
 
Although the proposed successive IF and IF schemes are developed based on lattice codes in this paper, it is worthwhile mentioning that  both schemes can be readily implemented via practical binary codes by simply extending the recent works~\cite{Chae:16,Ahn-Chae-Kim-Kim:21}. Additionally, although channels are assumed to be fixed during the communication block in this paper, the proposed schemes can be extended to practical environments in which channels vary within a codeword by modifying~\eqref{Q} as explained in~\cite{Chae:16,Ahn-Chae-Kim-Kim:21,Bakoury:15}.

 \subsubsection{Receiver complexity} \label{subsebsec:complexity}
Note that the symbol detection complexity of the proposed successive IF scheme and the proposed IF scheme depends on the set $d_{\text{stream}}$. Denote the collection of all feasible $d_{\text{stream}}$ satisfying the constraints \eqref{condition 1} and \eqref{condition 2} by $\mathcal{D}$.
Then, for all $d_{\text{stream}}\in\mathcal{D}$, the worst-case computational complexity of the IF filter operation~\eqref{IF_filter} is given by $O\left(M_{\text{R}}M_{\text{T}}^2+M_{\text{T}}^3\right)$~\cite{Chae:16, Seethaler:04,Liu:09} for  both the proposed successive IF scheme and the proposed IF scheme. In addition, the worst-case computational complexity of  the proposed IF scheme to find an appropriate integer matrix $\mathbf{A}_i$ using the LLL algorithm, i.e., the complexity of solving~\eqref{eq:Integer_matrix_search_channel_variation}, is given by $O\left(M^4\log M\right)$, where $M=\min\{M_{\text{T}},M_{\text{R}}\}$~\cite{Gan:09}. In a similar vein,  the worst-case computational complexity of solving~\eqref{eq:Integer_matrix_search_channel_variation_2} for the proposed successive IF scheme is given by $O\left(M^5\log M\right)$~\cite{Ahn-Chae-Kim-Kim:21} since the LLL algorithm needs to be applied at most $ M$ times. However, the required computational complexity to find an integer matrix is negligible considering the entire decoding process, because the LLL algorithm is required to be performed only $|\mathcal{D}|$  times over $n$ transmissions, where $|\mathcal{D}|$ denotes the cardinality of $\mathcal{D}$ and $n$ is much greater than $M_\text{T} $ and   $M_\text{R}$.

%$O\left(M_{\text{R}}M_{\text{T}}^3\log M_{\text{T}}\right)$

Therefore, the overall worst-case computational complexity of the proposed successive IF scheme and the proposed IF scheme is dominated by  $O\left(M_{\text{R}}M_{\text{T}}^2+M_{\text{T}}^3\right)$ per symbol, which is the same as that of MMSE receivers and MMSE-SIC receivers~\cite{Seethaler:04,Liu:09}.

% Hence, the overall computational complexity of the proposed successive IF and IF schemes is dominated by  $O\left(|\mathcal{D}|\left(M_{\text{R}}M_{\text{T}}^2+M_{\text{T}}^3\right)\right)$ per symbol, which is the same as that of MMSE and MMSE-SIC receivers. Here, $|\mathcal{D}|$ denote the cardinality of $\mathcal{D}$.

\section{Numerical Analysis and Discussions}
In this section, we numerically evaluate the achievable sum rates and  rate regions of the proposed  successive IF- and IF-based interference management schemes in various environments. 

\subsection{Simulation Environments and Evaluation Methodologies}

In simulation, we assume the Rician block fading channel model, that is, the channel matrix $\mathbf{H}_{i,j}$ in \eqref{eq:system model} is assumed to be given by
\begin{align}
\mathbf{H}_{i,j}=\sqrt{\alpha_{i,j}}\left(\sqrt{\frac{1}{K+1}}\mathbf{H}_{\text{NLoS}, i,j}+\sqrt{\frac{K}{K+1}}\mathbf{H}_{\text{LoS},i,j}\right),
\end{align}
where $\alpha_{i,j}\geq 0$ denotes the relative channel strength of $\mathbf{H}_{i,j}$, $\mathbf{H}_{\text{NLoS},i,j}$ and $\mathbf{H}_{\text{LoS},i,j}$ represent the non-line-of-sight (NLoS) and line-of-sight (LoS) channel matrices between transmitter $j$ and receiver $i$, respectively, and $K\geq 0$ is the Rician $K$ factor. 
We further assume that each coefficient in $\mathbf{H}_{\text{NLoS},i,j}$  follows $\mathcal{CN}(0,1)$ and $\mathbf{H}_{\text{LoS},i,j}$ is given by
\begin{align}
\mathbf{H}_{\text{LoS},i,j}=\beta_{i,j}\mathbf{q}_{M_\text{R}}(\theta_{i,j})\mathbf{q}_{M_\text{T}}^\ast(\phi_{i,j}),
\end{align}
where 
\begin{align}\label{linear array response}
\mathbf{q}_M(\theta)=\left[\begin{array}{cc}1\\ \exp(-\iota \pi \cos\theta) \\ \vdots \\ \exp(-\iota \pi (M-1) \cos\theta) \end{array}\right]\in \mathbb{C}^{M\times 1},
\end{align}
$\beta_{i,j}\sim \mathcal{CN}(0,1)$ denotes the short-term fading coefficient of the LoS path between transmitter $j$ and receiver $i$, $\theta_{i,j}$ and $\phi_{i,j}$ denote the angles of incidence of the LoS path on the receive antenna array of the receiver $j$ and the transmit antenna array of transmitter $i$, respectively, and $\iota^2=-1$.
We assume that uniform linear antenna arrays are used at transmitters and receivers. 
In addition, it is assumed that $\phi_{i,j}$ and $\theta_{i,j}$ follow $\text{Unif}[0,2\pi)$ for all $i,j\in\{1,2\}$. %It is assumed that each channel coefficient is independently and identically distributed (i.i.d.) drawn from a continuous distribution for each channel realization and fixed as the same value during communication. 
Due to the key hole effect~\cite{Keyhole02,Tse_wireless}, $\text{rank}(\mathbf{H}_{\text{LoS},i,j})=1$ for all $i,j\in\{1,2\}$,  and hence $\mathbf{H}_{i,j}$ becomes an ill-conditioned matrix as $K$ increases.

Note that when evaluating sum rates and rate regions, we consider the outage rate satisfying a certain outage probability~\cite{Tse_wireless}. For sum rate evaluation, assuming that the sum rate $R_{\text{sum}}=R_1+R_2$ is achievable by a scheme for a given channel realization, the outage probability for a target sum rate $R_\text{target,sum}$ is defined as
\begin{align}
p_{\text{outage}}(R_{\text{target,sum}})=\text{Pr}\left(R_{\text{sum}}<R_\text{target,sum}\right).\label{outage probability}
\end{align}
Then the $p$\% outage sum rate is defined as 
\begin{align}
R_{\text{outage,sum}}(p)=\sup\Big\{R_\text{target,sum}: p_{\text{outage}}(R_{\text{target,sum}})\leq \frac{p}{100}\Big\}.\label{outage rate}
\end{align}
We can generalize~\eqref{outage probability} and~\eqref{outage rate} for a rate pair as follows: 
\begin{align}
&p_{\text{outage}}(R_{\text{target},1},R_{\text{target},2})=\text{Pr}\left(R_1<R_{\text{target},1}~\text{and}~R_2<R_{\text{target},2}\right), \\
&(R_{\text{outage},1}(p),R_{\text{outage},2}(p))=\sup\Big\{(R_{\text{target},1},R_{\text{target},2}): p_{\text{outage}}(R_{\text{target},1},R_{\text{target},2})\leq \frac{p}{100}\Big\}.
\end{align}
Here, $(R_1, R_2)$ denotes an achievable rate pair for a given channel realization and $(R_{\text{target},1},R_{\text{target},2})$ denotes the target rate pair.

More specifically, for sum rate evaluation, an appropriate set $d_{\text{stream}}\in\mathcal{D}$ that maximizes the sum rate  is numerically optimized for each channel realization.  The outage sum rate is then evaluated over many channel realizations. On the other hand, for rate region evaluation, we calculate an achievable rate pair $(R_1, R_2)$ for every possible set $d_{\text{stream}}\in\mathcal{D}$ for each  channel realization. Then the outage rate pair is evaluated for each given set $d_{\text{stream}}$ over many channel realizations. Finally, a rate region of each scheme is obtained by the convex hull of the boundary points of the outage rate pairs.

\subsection{Numerical Results}

\subsubsection{Benchmark schemes}
As benchmark schemes, we consider the interference management scheme using joint ML decoding receivers and the interference management scheme using MMSE-SIC receivers explained in Section \ref{subsec:remarks}.  For the considered joint ML decoding, it is assumed that transmitters use i.i.d Gaussian codebooks and each receiver $i\in\{1,2\}$ exhaustively finds the set of stream vectors most likely sent by the transmitters based on its received signal vector $\bar{\mathbf{Y}}_i=[\begin{array}{ccccc}\bar{\mathbf{y}}_i(1) & \bar{\mathbf{y}}_i(2) &\cdots& \bar{\mathbf{y}}_i(n) \end{array}]\in \mathbb{R}^{2M_{R}\times n}$.

In order to numerically analyze the contribution of each technical component incorporated into  the proposed successive IF, we also consider the proposed successive IF with sending common streams only, i.e., $d_{[\text{c}],i}=d_i\leq \min \{2M_{\text{T}}, 2M_{\text{R}}\}$ and $d_{[\text{p}],i}=0$ for all $i\in\{1,2\}$, the proposed successive IF with sending private streams only, i.e., $d_{[\text{p}],i}=d_i\leq \min \{2M_{\text{T}}, 2M_{\text{R}}\}$ and $d_{[\text{c}],i}=0$  for all $i\in\{1,2\}$, and the IF with sending private streams only without any rank adaptation, i.e.,  $d_{[\text{p}],i}=d_i=\min \{2M_{\text{T}}, 2M_{\text{R}}\}$ and $d_{[\text{c}],i}=0$ for all $i\in\{1,2\}$.

\begin{remark}

\textcolor{black}{As discussed in Section \ref{subsebsec:complexity}, the symbol detection complexity of the proposed successive IF scheme and the proposed IF  scheme is determined according to the set $d_{\text{stream}}$, but only the worst case is considered here for ease of comparison. Then, except for the joint ML receiver, the worst-case computational complexity of all the considered receivers  is given by  $O\left(M_{\text{R}}M_{\text{T}}^2+M_{\text{T}}^3\right)$ per symbol, since in the worst case,  all schemes have the same effective channel size. On the other hand, in case of the joint ML receiver, the worst-case computational complexity  is given by $O(\frac{2^{nRM_\text{T}}}{n})$ per symbol  \cite{Zhan:14, LNIT, Legnain:13}, which is much larger than that of the other receivers. \hfill$\lozenge$}

\end{remark}

\subsubsection{No CSIT case}
\begin{figure}[t!]
    \centering
    \subfigure[$\alpha_{i,j}=1$, $K=0$.]{
                     \includegraphics[width=0.47\textwidth]{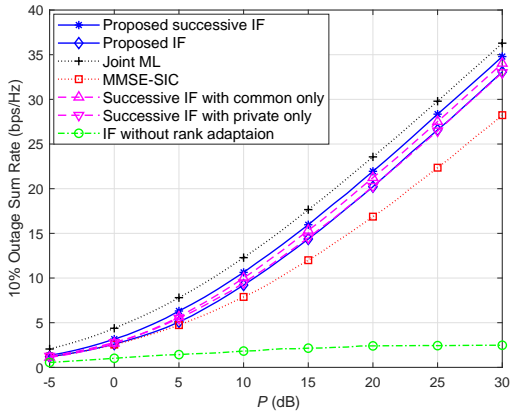}
                     \label{no CSIT:alpha_1, K_0}
    }
    \subfigure[$\alpha_{i,j}=1$, $K=20$.]{
                     \includegraphics[width=0.47\textwidth]{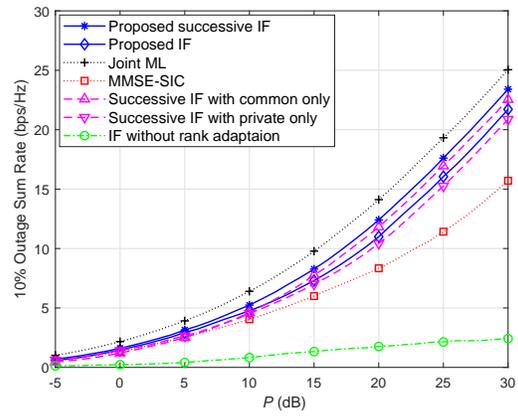}
                     \label{no CSIT:alpha_1, K_20}
    }
    \\
    \subfigure[$\alpha_{i,j}=0.25$, $K=0$.]{
                     \includegraphics[width=0.47\textwidth]{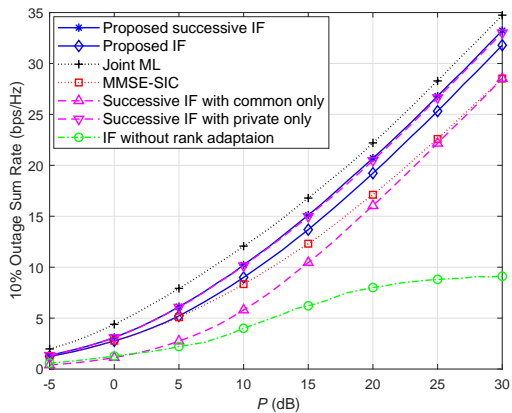}
                     \label{no CSIT:alpha_0.25, K_0}
    }
    \subfigure[$\alpha_{i,j}=0.25$, $K=20$.]{
                     \includegraphics[width=0.47\textwidth]{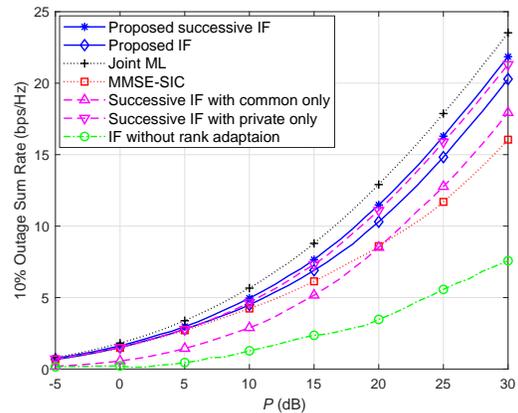}
                     \label{no CSIT:alpha_0.25, K_20}
    }
    \caption{Achievable  outage sum rates of the no CSIT case when $M_\text{T}=8$, $M_\text{R}=4$, and $\alpha_{i,i}=1$, $\forall i=1,2$.}
    \label{FIG:no CSIT_sum rate}

\end{figure}

\begin{figure}[t!]
    \centering
    \subfigure[$\alpha_{i,j}=1$, $K=0$.]{
                     \includegraphics[width=0.47\textwidth]{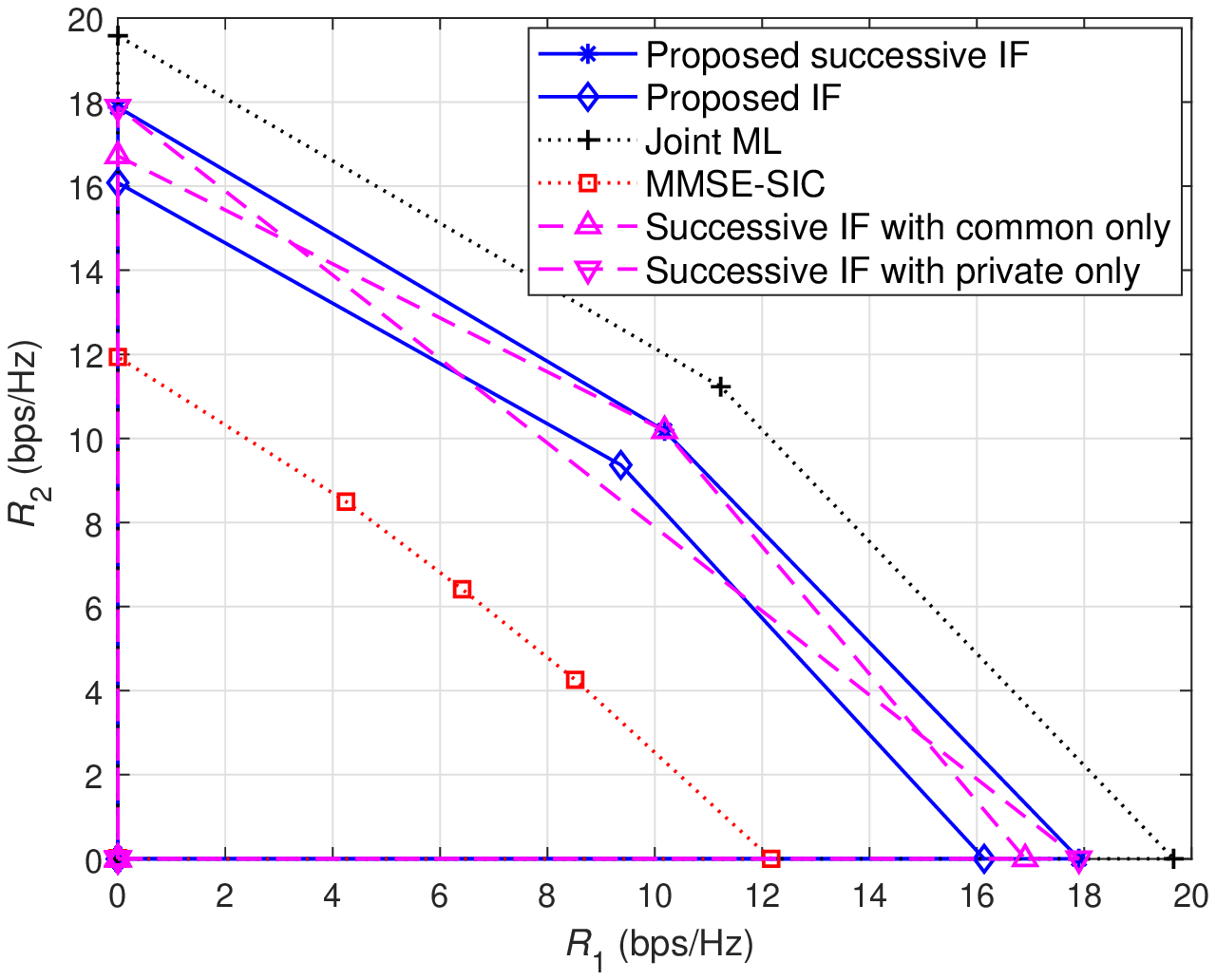}
                     \label{no CSIT:alpha_1, K_0}
    }
    \subfigure[$\alpha_{i,j}=1$, $K=20$.]{
                     \includegraphics[width=0.47\textwidth]{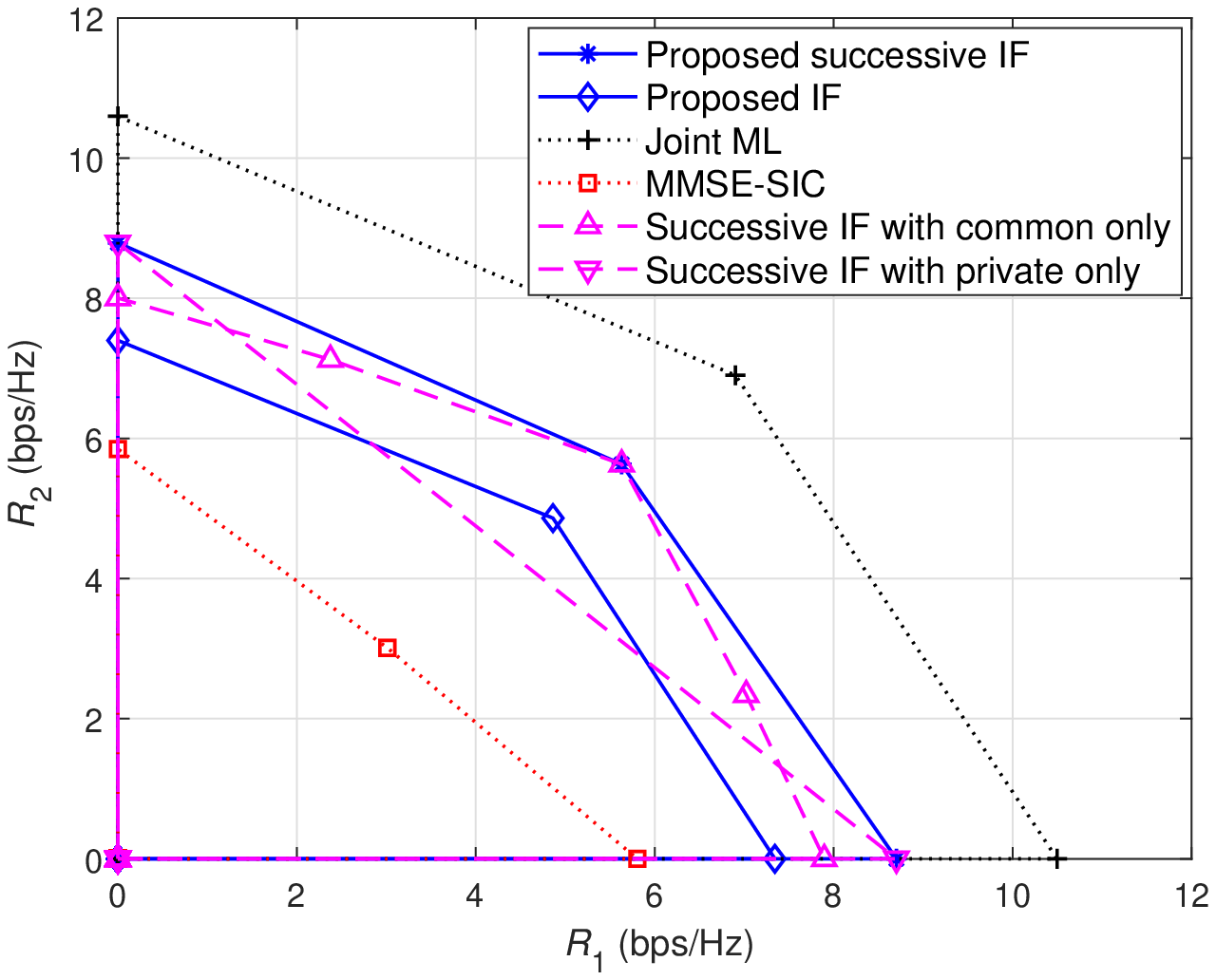}
                     \label{no CSIT:alpha_1, K_20}
    }
    \\
    \subfigure[$\alpha_{i,j}=0.25$, $K=0$.]{
                     \includegraphics[width=0.47\textwidth]{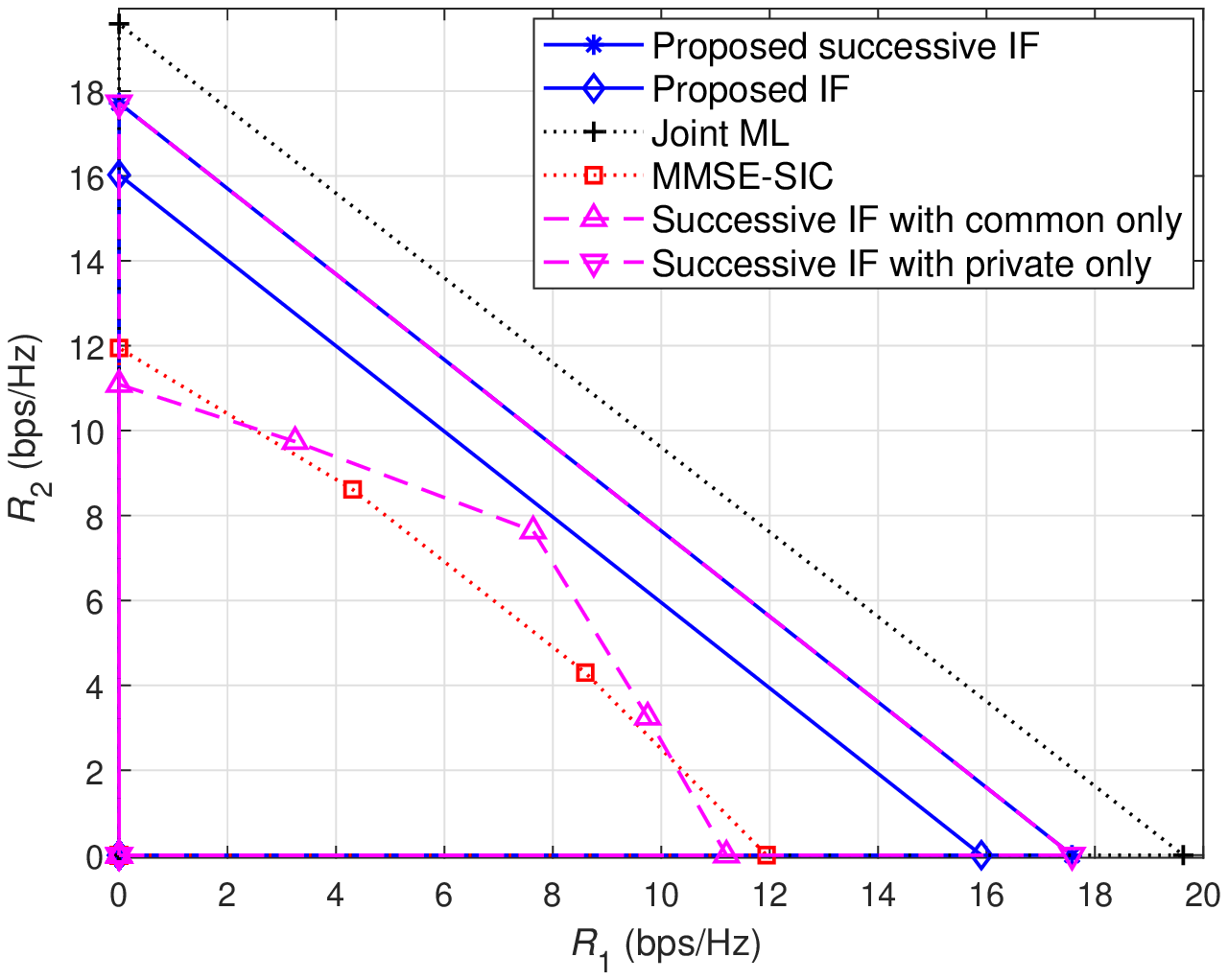}
                     \label{no CSIT:alpha_0.25, K_0}
    }
    \subfigure[$\alpha_{i,j}=0.25$, $K=20$.]{
                     \includegraphics[width=0.47\textwidth]{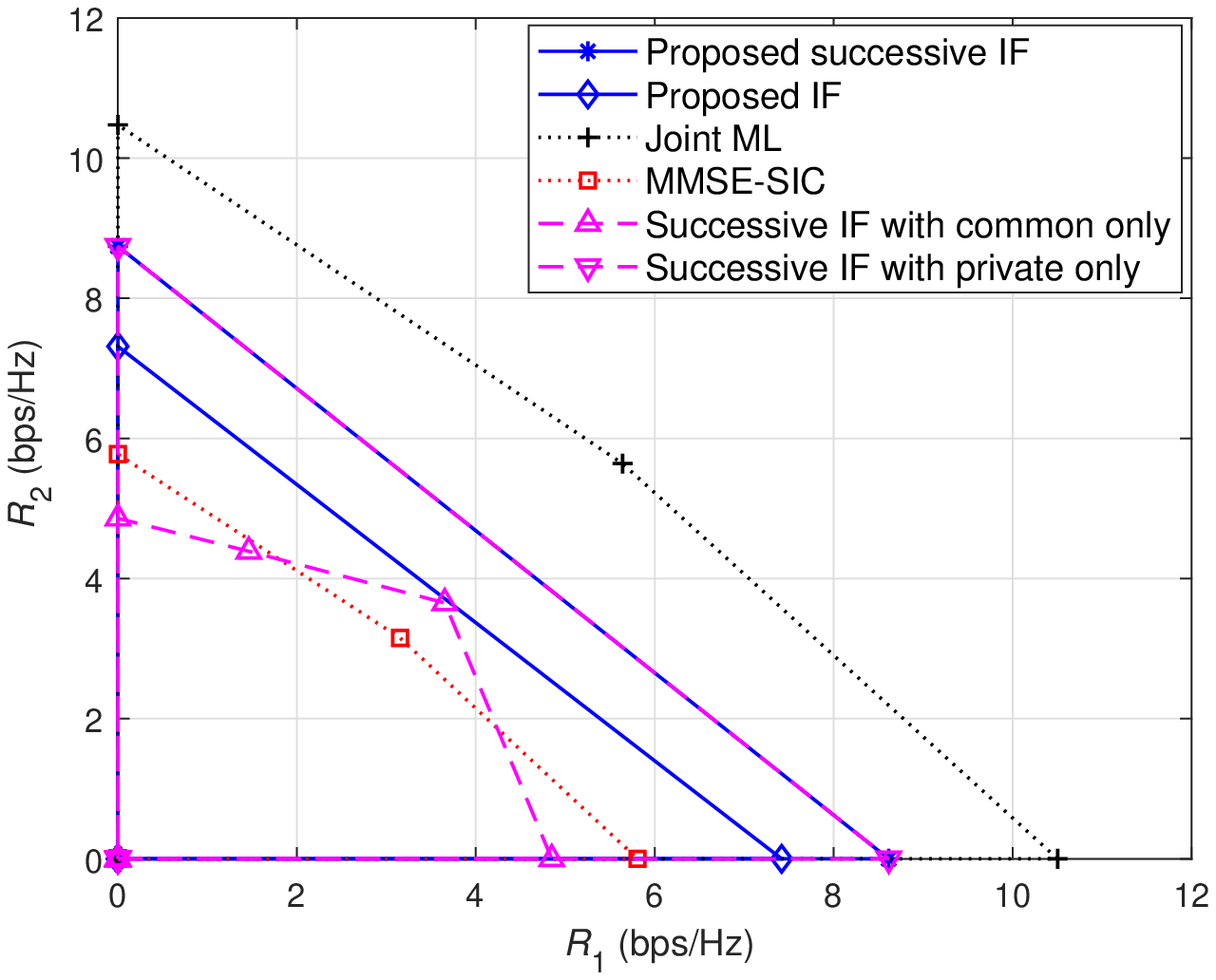}
                     \label{no CSIT:alpha_0.25, K_20}
    }
    \caption{Achievable outage rate regions of the no CSIT case when $M_\text{T}=8$, $M_\text{R}=4$, $P=20\text{dB}$, and $\alpha_{i,i}=1$, $\forall i=1,2$.}
    \label{FIG:no CSIT_rate region}

\end{figure}

First, we consider the no CSIT case. Recall that the transmit beamforming matrices are set to~\eqref{no CSIT} by utilizing an identity matrix due to the lack of CSIT. 
In simulation, we set $M_\text{T}=8$, $M_\text{R}=4$, $\alpha_{1,1}=\alpha_{2,2}=1$, $\alpha_{1,2}, \alpha_{2,1}\in\{0.25, 1\}$, and $K\in\{0, 20\}$. \textcolor{black}{To examine the effect of the channel strength of interfering links on the achievable rate, we consider both the relatively strong interference case ($\alpha_{1,2}=\alpha_{2,1}=1$) and the relatively weak interference case ($\alpha_{1,2}=\alpha_{2,1}=0.25$). Additionally, we also consider both cases where the LoS components are relatively dominant ($K=10$) or non-existent ($K=0$) to see the effect of channel condition number on the achievable rate.} Under this setting, we plot the achievable $10\%$ outage sum rates and rate regions of the considered schemes in Fig.~\ref{FIG:no CSIT_sum rate} and Fig.~\ref{FIG:no CSIT_rate region}, respectively.

%For comparison, we also plot the outage sum rate of the IF scheme with sending private streams only without any rank adaptation, i.e., the previous interference management scheme~\cite[Section VII-C]{Zhan:14} explained in Remark~\ref{Remark1}. 

The results demonstrate that both the proposed successive IF scheme and the proposed IF scheme strictly outperform the MMSE-SIC scheme. In particular, the performance gap increases with increasing $K$ due to the fact that the achievable rate of the MMSE-SIC scheme is  significantly degraded compared to that of the successive IF scheme or the IF scheme for ill-conditioned channels, since in case of the MMSE-SIC scheme, severe noise amplification is caused by restricting $\mathbf{A}_{\text{SIC},i}$ to
an identity matrix for all $i\in\{1,2\}$. Figs.~\ref{FIG:no CSIT_sum rate} and~\ref{FIG:no CSIT_rate region} also show that the gap between the achievable rates of the joint ML decoding and the proposed successive IF is small. 
Therefore, it can be seen that compared to the joint ML decoding receiver, the proposed successive IF receiver and the proposed IF receiver can significantly reduce the receiver complexity while only slightly lowering achievable rates. See Section \ref{subsebsec:complexity} and Remark 2 for more detailed analysis of receiver complexity.

In Fig.~\ref{FIG:no CSIT_sum rate}, it is also observed that a significant rate improvement can be obtained from rank adaptation and/or message splitting (employing both common and private streams), by comparing the proposed IF scheme with the case of sending private streams only without rank adaptation, i.e., $d_{[\text{p}],i}=d_i= \min \{2M_{\text{T}}, 2M_{\text{R}}\}$ and $d_{[\text{c}],i}=0$  for all $i\in\{1,2\}$. As seen in the figure, rank adaptation is essentially required to improve achievable rates. Hence, for the rest of the performance comparison in this section, we omit the case of no rank adaptation.

In addition,  by comparing the achievable outage rate of the proposed successive IF scheme with that of the scheme with sending common or private streams only, it is shown that employing common streams becomes more beneficial as  $\alpha_{1,2}$ and $\alpha_{2,1}$ increase, since common streams are required to be decoded by both receivers. More specifically, when $\alpha_{1,2}=\alpha_{2,1}=1$, sending only common streams  becomes a near-optimal strategy in terms of maximizing the sum rate under the proposed scheme in a similar vein to the results of previous works~\cite{Costa87,Sato;81}. On the other hand, when $\alpha_{1,2}=\alpha_{2,1}=0.25$, sending only private streams becomes a near-optimal strategy.

Moreover, interestingly,  the results demonstrate that employing common streams becomes more beneficial as $K$ increases. Note that channel matrices become ill-conditioned when $K$ is large, resulting in an overall decrease in the proper number of transmitted streams. Hence, in this case, each receiver can more easily recover all common streams sent by both transmitters, since the total number of streams to be decoded by each receiver becomes small.

%
%In addition, it is observed that a significant rate improvement can be obtained from rank adaptation and/or message splitting (employing both common and private streams), especially when the interfering channel strength is strong, i.e., when $\alpha_{i,j}$ is large. By comparing the achievable rate of the proposed successive IF with that of the scheme with sending common or private streams only, it is also shown that employing common streams becomes more beneficial as  $\alpha_{i,j}$ increases due to the fact that common streams are required to be decoded by both receivers. More specifically, when $\alpha_{i,j}=1$, sending only common streams  becomes a near-optimal strategy under the proposed scheme as shown in Figs.~\ref{no CSIT:alpha_1, K_0} and \ref{no CSIT:alpha_1, K_20}, similar to the results of previous works~\cite{Costa87,Sato;81}. On the other hand, when $\alpha_{i,j}=0.25$ and $K=0$, sending only private streams is a near-optimal strategy as shown in Fig~\ref{no CSIT:alpha_0.25, K_0}. 

\begin{figure}[t!]
    \centering
    \subfigure[$\alpha_{i,j}=1$, $K=0$.]{
                     \includegraphics[width=0.47\textwidth]{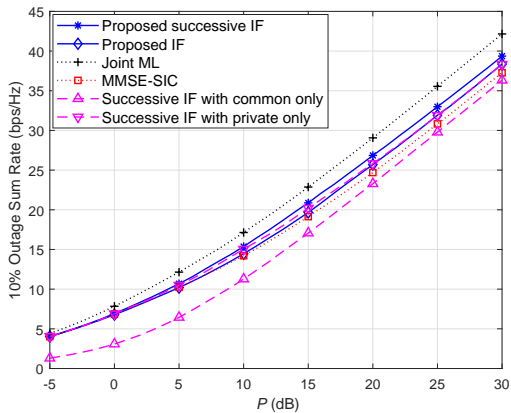}
                     \label{partial CSIT:alpha_1, K_0}
    }
    \subfigure[$\alpha_{i,j}=1$, $K=20$.]{
                     \includegraphics[width=0.47\textwidth]{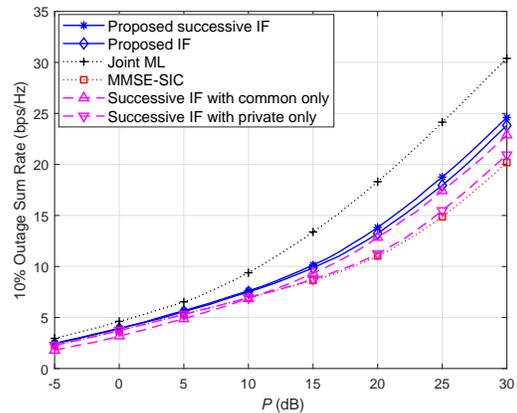}
                     \label{partial CSIT:alpha_1, K_20}
    }
    \\
    \subfigure[$\alpha_{i,j}=0.25$, $K=0$.]{
                     \includegraphics[width=0.47\textwidth]{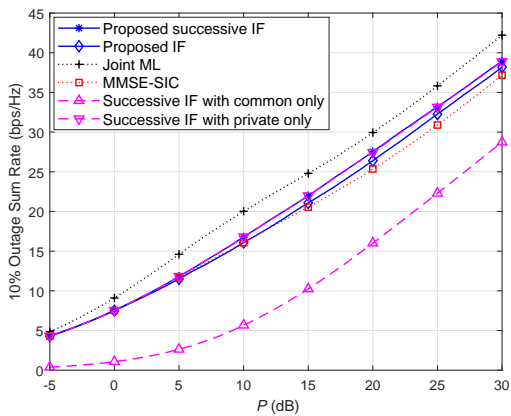}
                     \label{partial CSIT:alpha_0.25, K_0}
    }
    \subfigure[$\alpha_{i,j}=0.25$, $K=20$.]{
                     \includegraphics[width=0.47\textwidth]{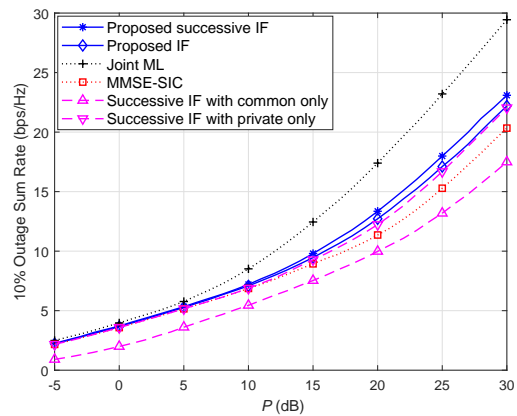}
                     \label{partial CSIT:alpha_0.25, K_20}
    }
    \caption{Achievable outage sum rates of the partial CSIT case when $M_\text{T}=8$, $M_\text{R}=4$, and $\alpha_{i,i}=1$, $\forall i=1,2$.}
    \label{FIG:partial CSIT_sum rate}

\end{figure}

\subsubsection{Partial CSIT case}
\begin{figure}[t!]
    \centering
    \subfigure[$\alpha_{i,j}=1$, $K=0$.]{
                     \includegraphics[width=0.47\textwidth]{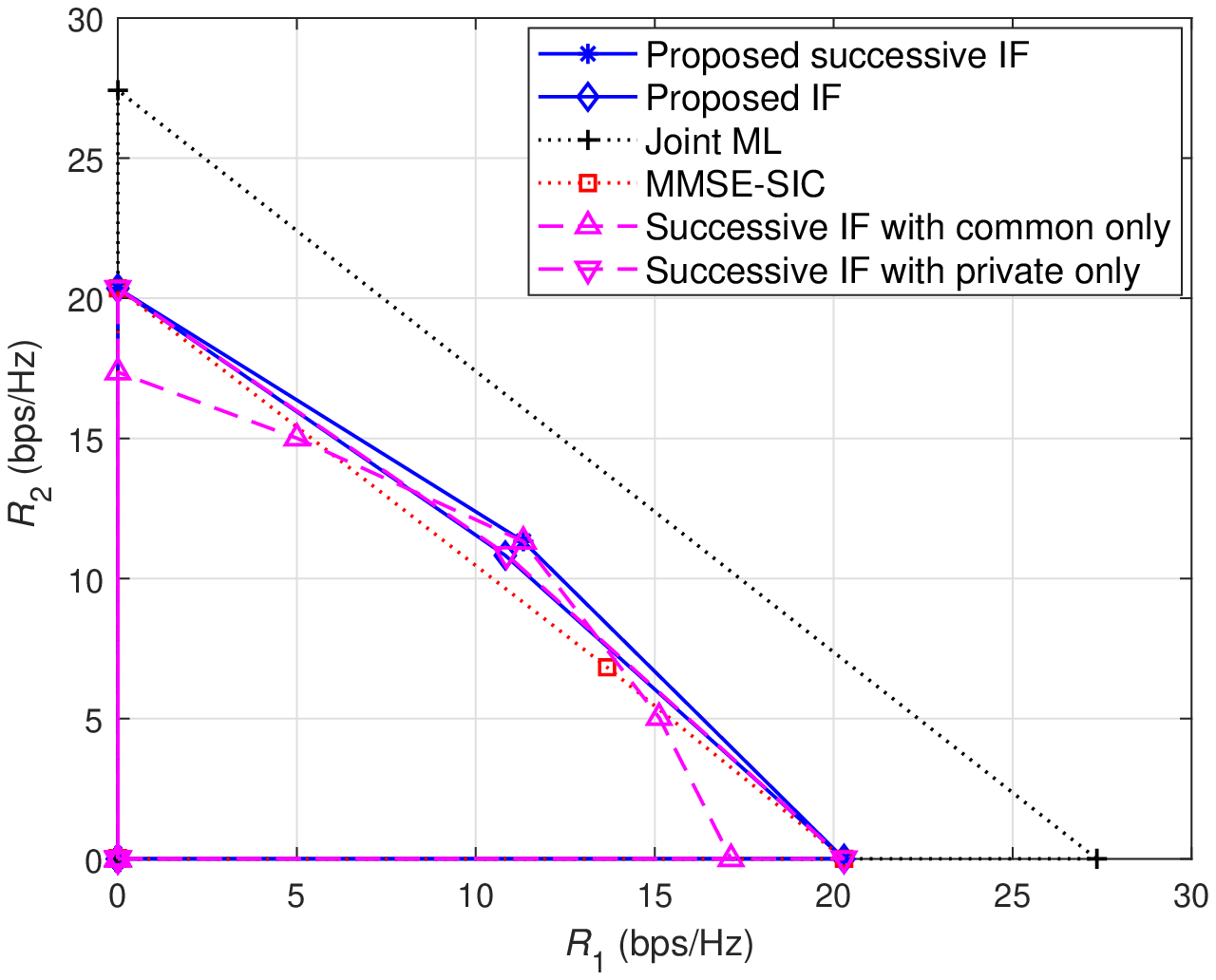}
                     \label{partial CSIT:alpha_1, K_0}
    }
    \subfigure[$\alpha_{i,j}=1$, $K=20$.]{
                     \includegraphics[width=0.47\textwidth]{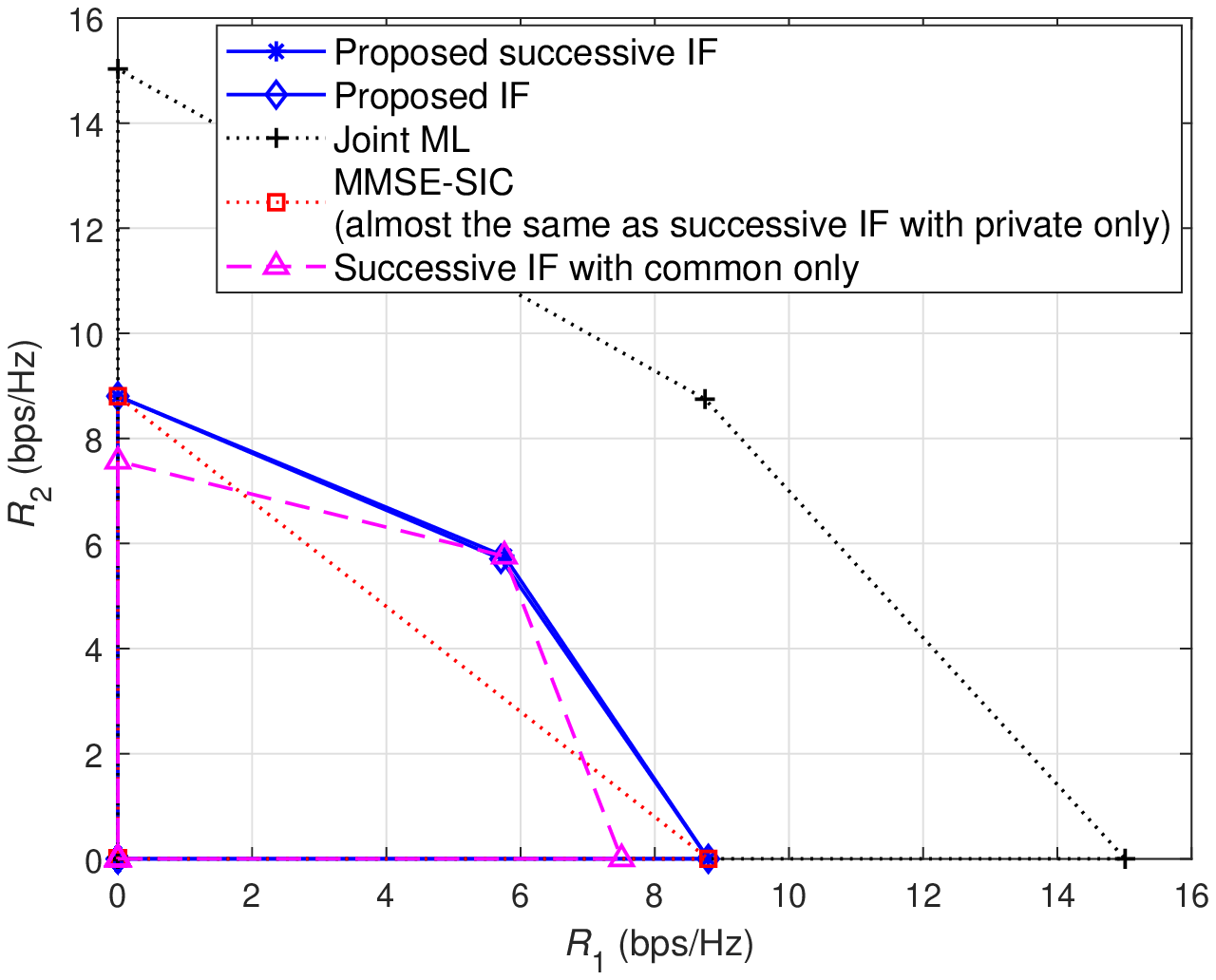}
                     \label{partial CSIT:alpha_1, K_20}
    }
    \\
    \subfigure[$\alpha_{i,j}=0.25$, $K=0$.]{
                     \includegraphics[width=0.47\textwidth]{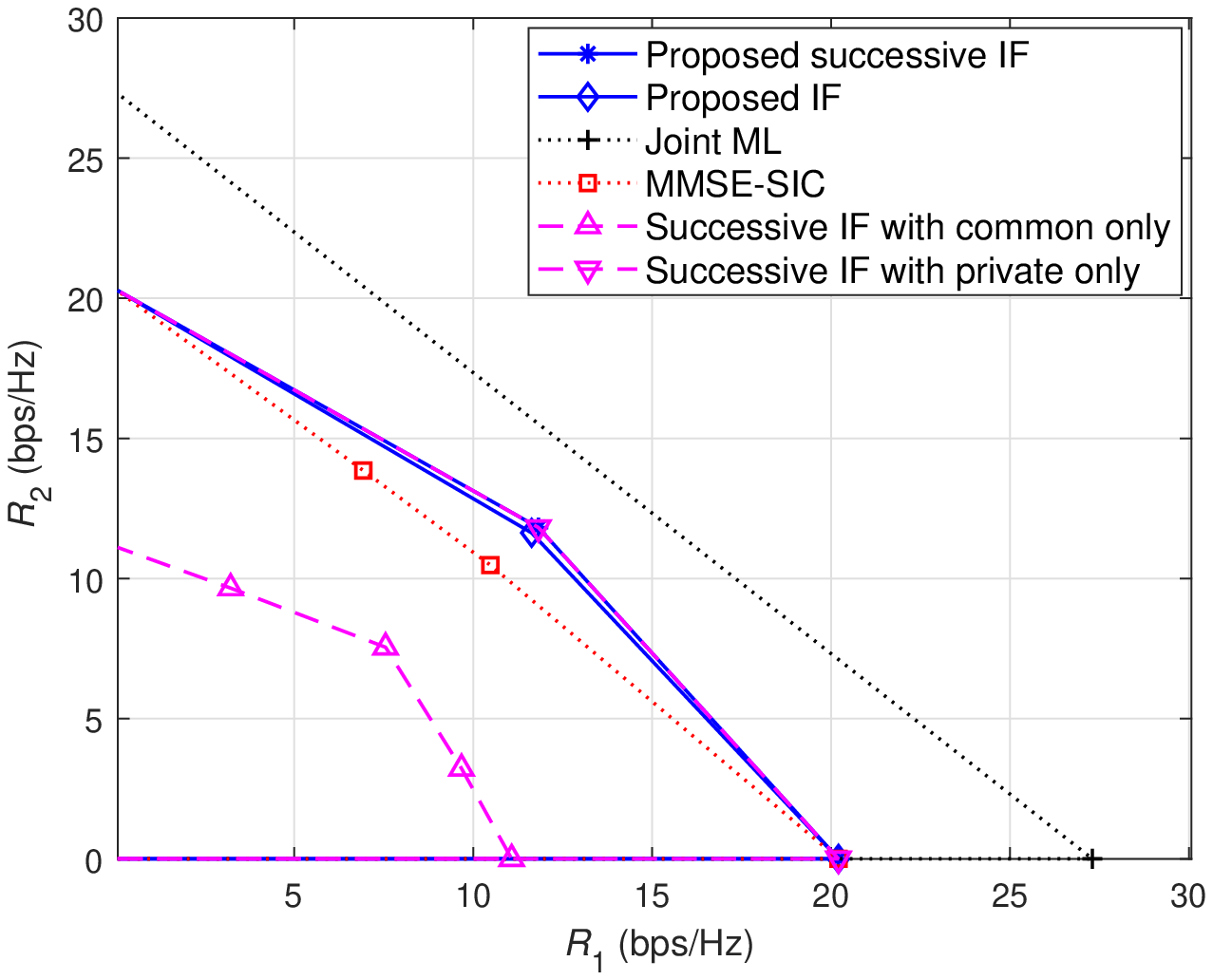}
                     \label{partial CSIT:alpha_0.25, K_0}
    }
    \subfigure[$\alpha_{i,j}=0.25$, $K=20$.]{
                     \includegraphics[width=0.47\textwidth]{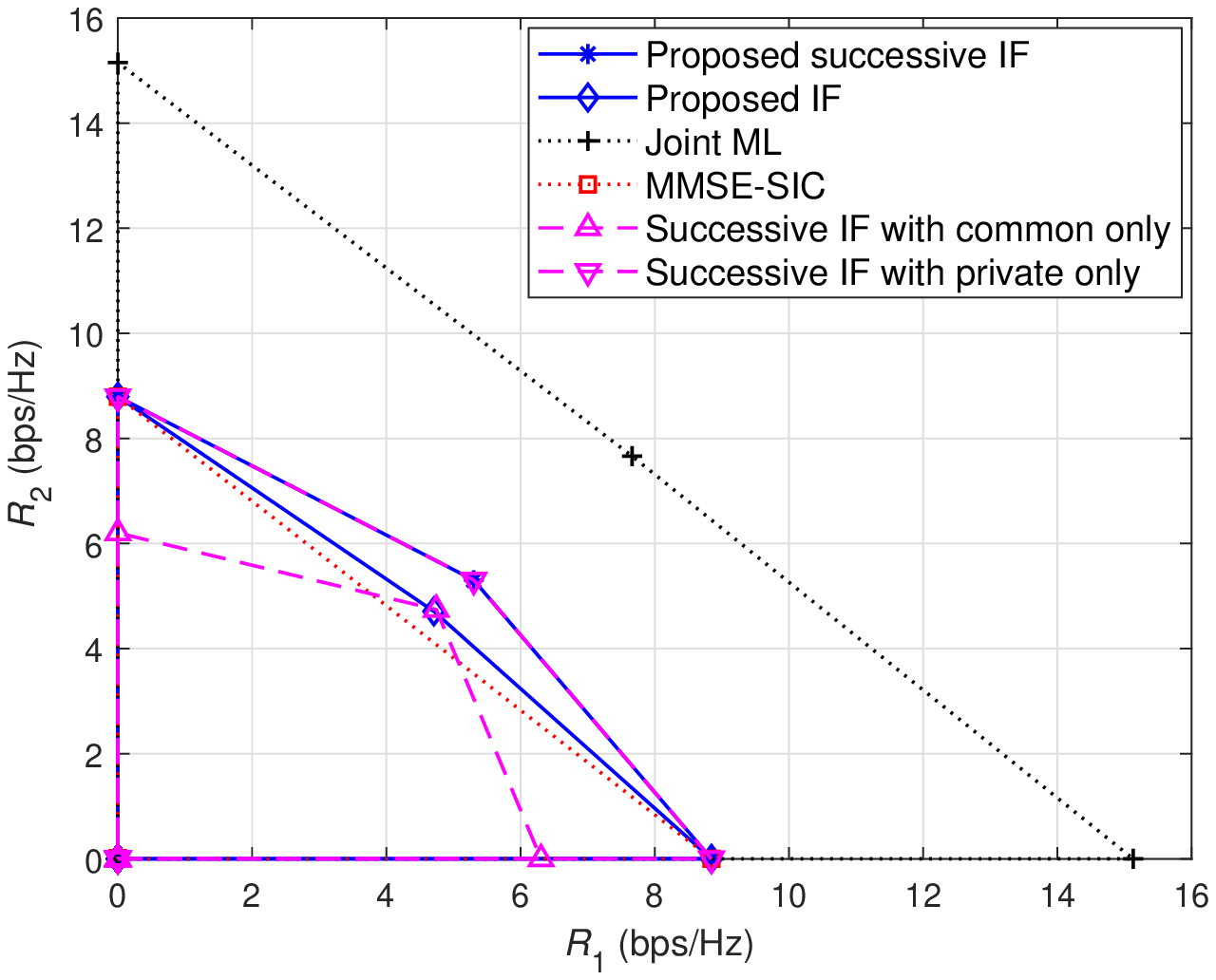}
                     \label{partial CSIT:alpha_0.25, K_20}
    }
    \caption{Achievable outage rate regions of the partial CSIT case when $M_\text{T}=8$, $M_\text{R}=4$, $P=20\text{dB}$, and $\alpha_{i,i}=1$, $\forall i=1,2$.}
    \label{FIG:partial CSIT_rate region}

\end{figure}

Next, we  perform numerical simulation of the partial CSIT case.  Recall that in this case, the transmit beamforming matrices are set to~\eqref{partial CSIT} by utilizing the CSI of  $\bar{\mathbf{H}}_{i,i}$. Except for the CSIT assumption, other simulation parameters are the same as in the simulation of the no CSIT case explained above. The achievable $10\%$ outage sum rates and rate regions of the considered schemes for the partial CSIT case are plotted in Figs.~\ref{FIG:partial CSIT_sum rate} and~\ref{FIG:partial CSIT_rate region}, respectively.

By comparing Figs.~\ref{FIG:partial CSIT_sum rate} and~\ref{FIG:partial CSIT_rate region} with  Figs.~\ref{FIG:no CSIT_sum rate} and~\ref{FIG:no CSIT_rate region}, it is shown that the overall achievable outage rates of the partial CSIT case increase compared to the no CSIT case thanks to the presence of CSIT of direct links. Specifically, since the singular values of the channel matrix of the desired link are preserved in the effective channel matrix obtained after applying transmit beamforming if the SVD-based beamforming~\eqref{partial CSIT} is employed, the use of private streams has been shown to be more advantageous for the partial CSIT case compared to the no CSIT case that simply uses an identity matrix as a transmit beamforming matrix. Additionally, it is observed that the outage rate performance tendencies with respect to $\alpha_{i,j}$ and $K$ are similar to those without CSIT.

\begin{figure}[t!]
    \centering
    \subfigure[$M_{\text{T}}=6$, $K=0$.]{
                     \includegraphics[width=0.47\textwidth]{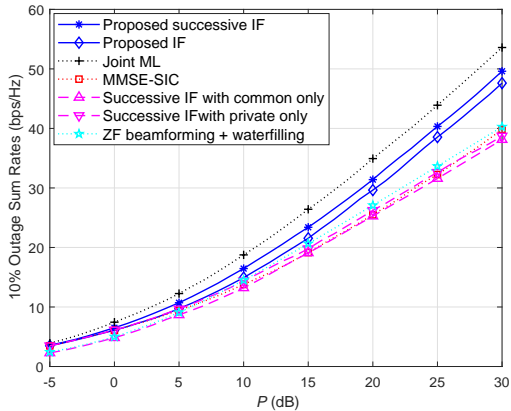}
                     \label{M=6, K=0}
    }
    \subfigure[$M_{\text{T}}=8$, $K=0$.]{
                     \includegraphics[width=0.47\textwidth]{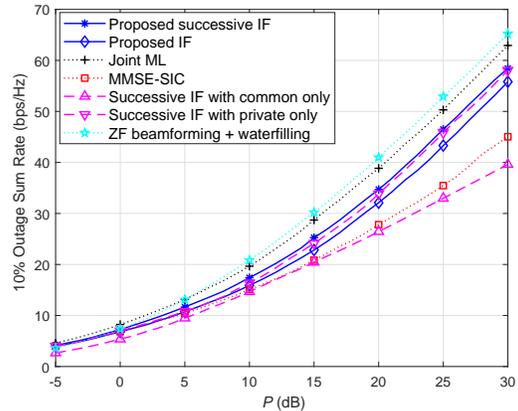}
                     \label{M=8, K=0}
    }
    \subfigure[$M_{\text{T}}=6$, $K=20$.]{
                     \includegraphics[width=0.47\textwidth]{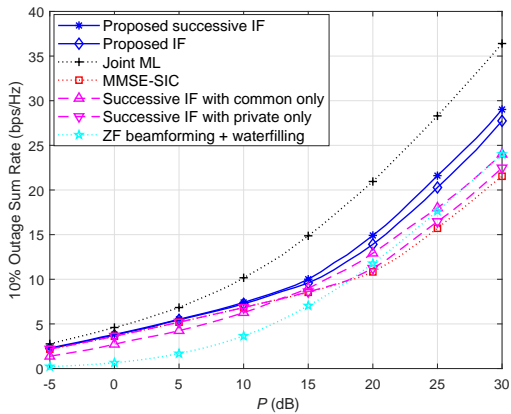}
                     \label{M=6, K=20}
    }
    \subfigure[$M_{\text{T}}=8$, $K=20$.]{
                     \includegraphics[width=0.47\textwidth]{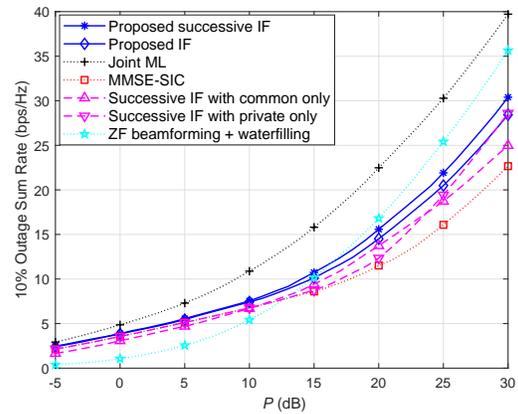}
                     \label{M=8, K=20}
    }
    \caption{Achievable outage sum rates of the full CSIT case when $M_\text{R}=4$ and $\alpha_{i,i}=\alpha_{i,j}=1$, $\forall i,j=1,2$ where $i\neq j$.}
    \label{FIG:full CSIT_sum rate}

\end{figure}

\subsubsection{Full CSIT case}
\begin{figure}[t!]
    \centering
    \subfigure[$M_{\text{T}}=6$, $K=0$.]{
                     \includegraphics[width=0.47\textwidth]{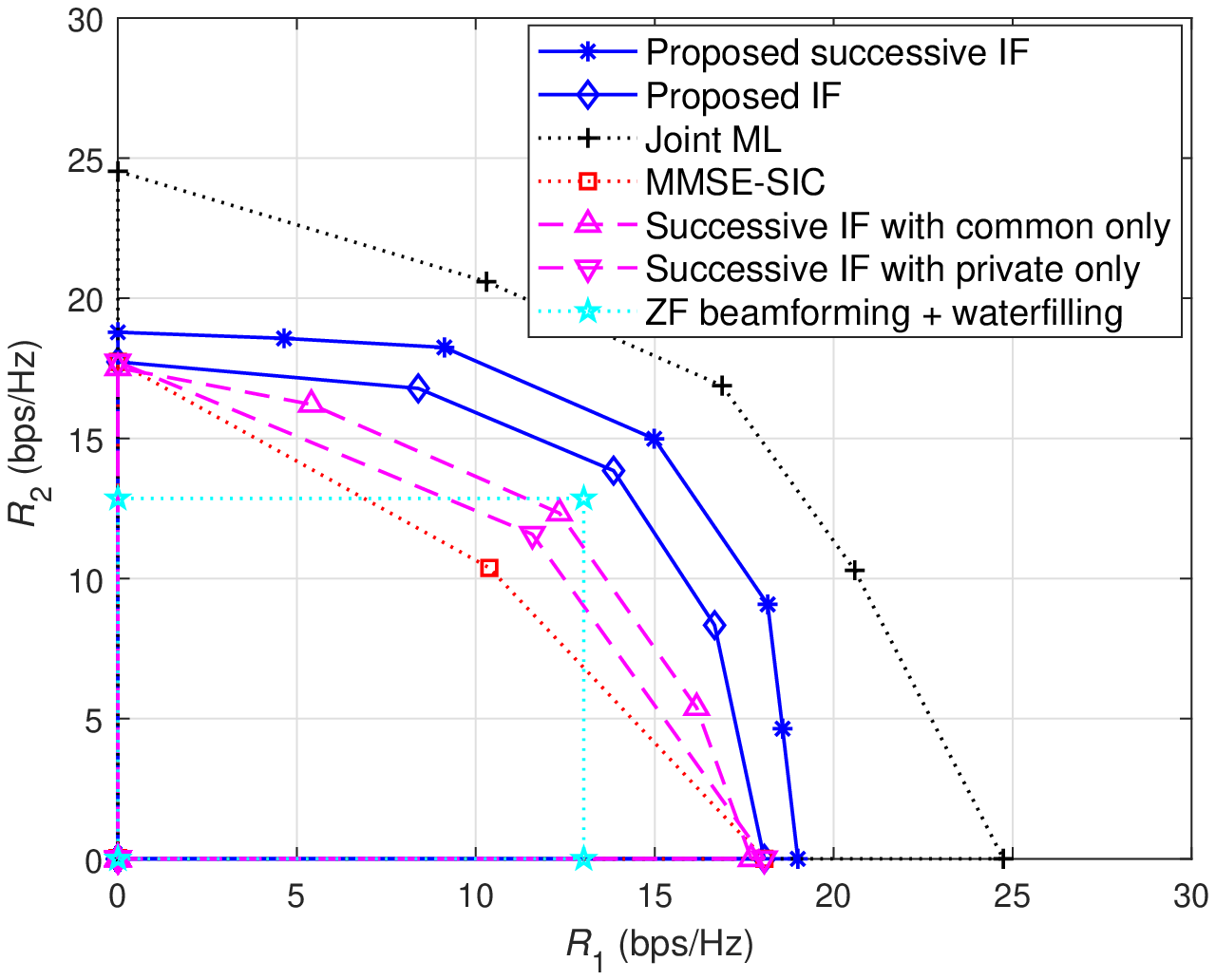}
                     \label{M=6, K=0}
    }
    \subfigure[$M_{\text{T}}=8$, $K=0$.]{
                     \includegraphics[width=0.47\textwidth]{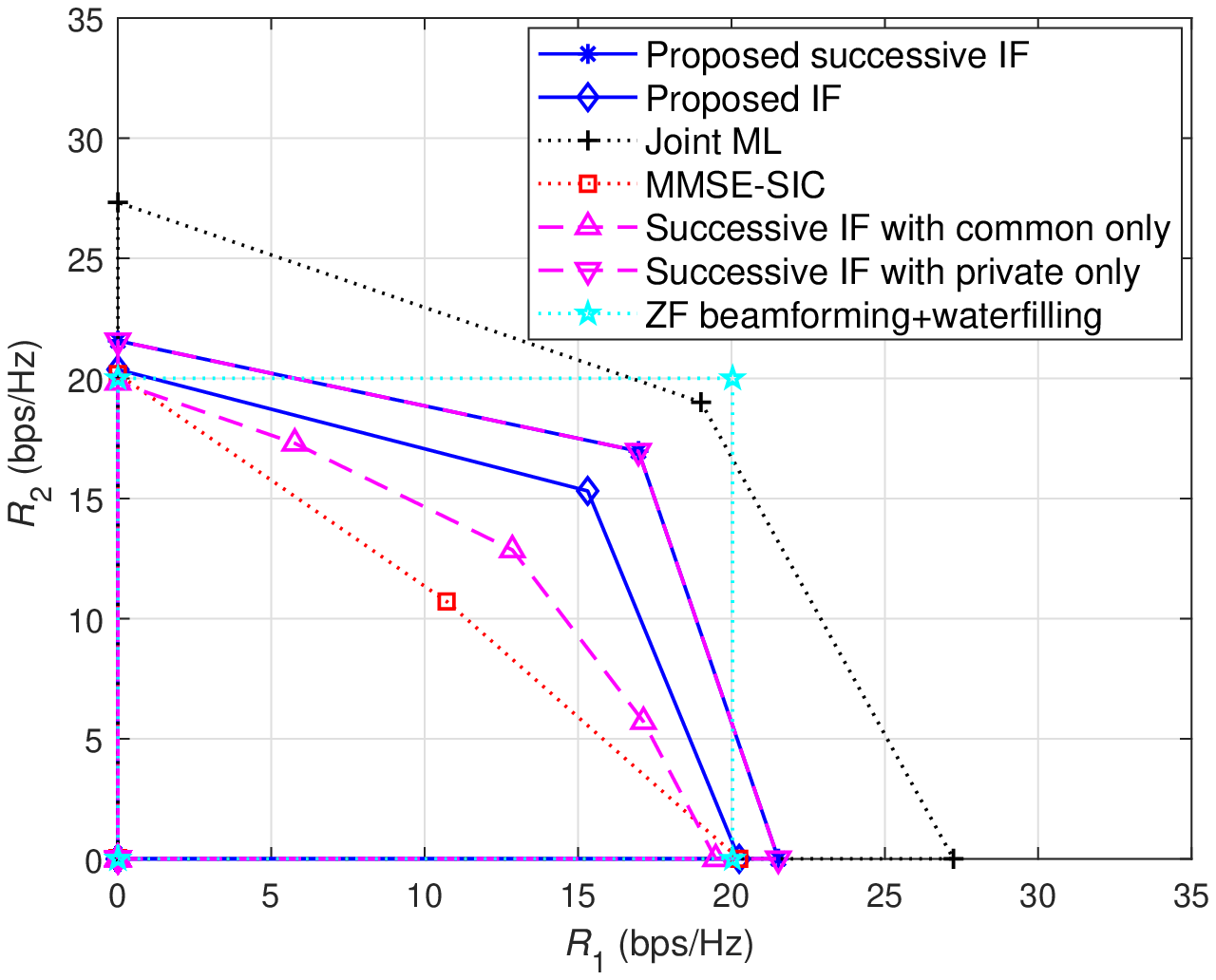}
                     \label{M=8, K=0}
    }
    \subfigure[$M_{\text{T}}=6$, $K=20$.]{
                     \includegraphics[width=0.47\textwidth]{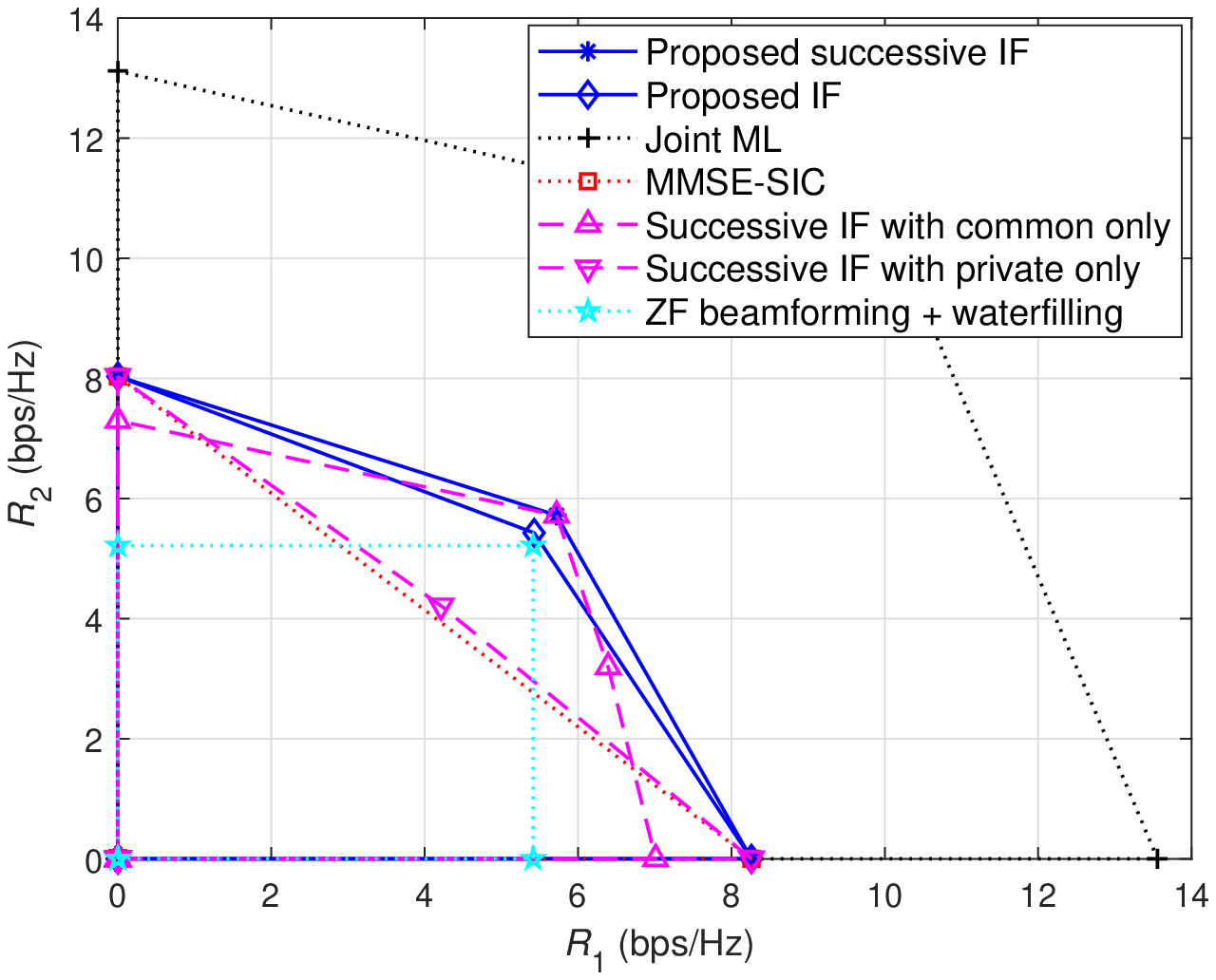}
                     \label{M=6, K=20}
    }
    \subfigure[$M_{\text{T}}=8$, $K=20$.]{
                     \includegraphics[width=0.47\textwidth]{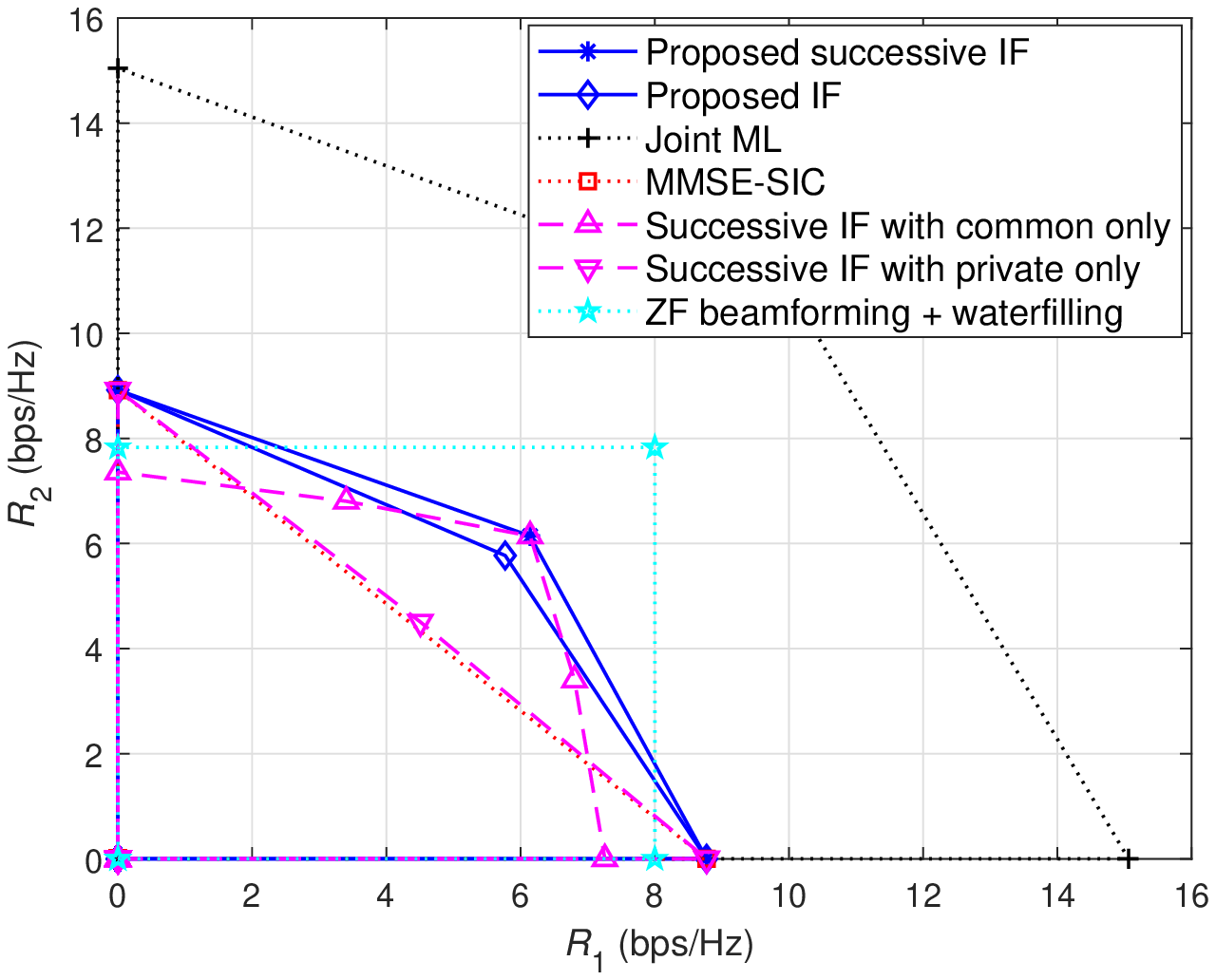}
                     \label{M=8, K=20}
    }
    \caption{Achievable outage rate regions of the full CSIT case when $M_\text{R}=4$ and $\alpha_{i,i}=\alpha_{i,j}=1$, $\forall i,j=1,2$ where $i\neq j$.}
    \label{FIG:full CSIT_rate region}

\end{figure}

Finally, we examine the full CSIT case. Recall that in this case, the transmit beamforming matrices for common and private streams are set to~\eqref{full CSIT-common} and~\eqref{full CSIT-private}, respectively. In simulation, we set $M_\text{R}=4$ and $\alpha_{i,j}=1$ for all $i,j\in\{1,2\}$, and consider the cases where $M_\text{T}\in\{6,8 \}$ and $K\in\{0,20\}$. In addition, we consider a transmission scheme using ZF beamforming and water-filling power allocation as a conventional benchmark scheme. More specifically, in this benchmark scheme, by utilizing the CSIT of cross links, inter-user interference is completely removed by ZF beamforming of each transmitter. Each transmitter--receiver pair then performs SVD beamforming and water-filling power allocation based on its acquired interference-free single-user channel, which can achieve the capacity of the given single-user channel. Under this setting, the achievable $10\%$ outage sum rates and rate regions of the considered schemes are depicted in Fig.~\ref{FIG:full CSIT_sum rate} and  Fig.~\ref{FIG:full CSIT_rate region}, respectively.

From Figs.~\ref{FIG:full CSIT_sum rate} and~\ref{FIG:full CSIT_rate region}, it can be seen that the overall achievable outage rates of the full CSIT case are further improved compared to the no CSIT case (Figs.~\ref{FIG:no CSIT_sum rate} and~\ref{FIG:no CSIT_rate region})
and the partial CSIT case (Figs.~\ref{FIG:partial CSIT_sum rate} and~\ref{FIG:partial CSIT_rate region}) thanks to the presence of CSIT of both direct links and cross links. Specifically, when $M_\text{T}$ is large enough, i.e., $M_\text{T}=8$, all interference links can be completely nulled out by transmit ZF beamforming without reducing the maximum number of streams that can be transmitted from each transmitter. Hence, when $K=0$, i.e., for well-conditioned channels, it is observed that sending private streams only achieves a near-optimal outage sum rate under the proposed scheme, as shown in Fig.~\ref{M=8, K=0}. On the other hand, even when $M_\text{T}$ is large, if the channels are highly ill-conditioned, i.e., $K=20$, the results demonstrate that the proposed message splitting method achieves better performance compared to the case with sending private streams only due to the gain from employing common streams as shown in Fig.~\ref{M=8, K=20}, as similar to the no CSIT case and the partial CSIT case explained above. This tendency becomes more significant as $M_\text{T}$ is smaller, because the dimension of null space of a cross link used to send private streams becomes smaller. 

Moreover, it is also observed in Figs.~\ref{M=6, K=0} and~\ref{M=6, K=20} that the achievable outage rate of the benchmark scheme using ZF beamforming and water-filling power allocation considerably decreases when $M_\text{T}$ is small as $M_\text{T}=6$, because the rank of the interference-free single-user channel matrix obtained after applying  transmit ZF beamforming becomes only two. As a result,  it can be seen that the proposed scheme combining message splitting and IF sum decoding can strictly outperform the benchmark scheme especially when $M_\text{T}$ is small, although equal power allocation is applied for all streams in the proposed scheme.\footnote{Note that if we restrict streams to be allocated  equal power in the benchmark scheme, it is easy to see that the joint ML decoding always outperforms the benchmark scheme.}

% for the case where $M_\text{T}=8$, 

\section{Conclusion}
In this paper, we have developed a new interference management scheme based on IF and message splitting for the two-user MIMO interference channel. Combining IF sum decoding and a message-splitting method  that uses both common and private streams, the proposed scheme achieves good performance while effectively managing interference with low complexity. In addition, we considered three representative CSIT assumptions widely adopted in the literature and proposed low complexity transmit beamforming suitable for each CSIT assumption. Considering the performance gain confirmed through numerical simulations and the fact that the proposed scheme can be easily implemented with practical binary codes, our work could be one of the promising techniques for interference management in beyond 5G and 6G communication systems.

Moreover, the ideas of  our work can be extended in several interesting research directions. For example, extending to more than three user cases and combining the proposed scheme with interference alignment would be an interesting topic.

\bibliographystyle{IEEEtran}
\bibliography{IEEEabrv,References}

\begin{thebibliography}{10}
\providecommand{\url}[1]{#1}
\csname url@samestyle\endcsname
\providecommand{\newblock}{\relax}
\providecommand{\bibinfo}[2]{#2}
\providecommand{\BIBentrySTDinterwordspacing}{\spaceskip=0pt\relax}
\providecommand{\BIBentryALTinterwordstretchfactor}{4}
\providecommand{\BIBentryALTinterwordspacing}{\spaceskip=\fontdimen2\font plus
\BIBentryALTinterwordstretchfactor\fontdimen3\font minus
  \fontdimen4\font\relax}
\providecommand{\BIBforeignlanguage}[2]{{%
\expandafter\ifx\csname l@#1\endcsname\relax
\typeout{** WARNING: IEEEtran.bst: No hyphenation pattern has been}%
\typeout{** loaded for the language `#1'. Using the pattern for}%
\typeout{** the default language instead.}%
\else
\language=\csname l@#1\endcsname
\fi
#2}}
\providecommand{\BIBdecl}{\relax}
\BIBdecl

\bibitem{Samsung:20}
{Samsung Research}, ``{6G white paper: The next hyper connected experience for
  all},'' pp. 1--46, [Online]. Available:
  \text{https://cdn.codeground.org/nsr/downloads/researchareas/6G\%20Vision.pdf},
  {Dec. 2020.}

\bibitem{Dang:20}
S.~Dang, O.~Amin, B.~Shihada, and M.-S. Alouini, ``{What should 6G be?}''
  \emph{Nat. Electron.}, no.~3, pp. 20--29, Jan. 2020.

\bibitem{Giordani:20}
M.~Giordani, M.~Polese, M.~Mezzavilla, S.~Rangan, and M.~Zorzi, ``Toward {6G}
  networks: {U}se cases and technologies,'' \emph{{IEEE} Commun. Mag.},
  vol.~58, no.~3, pp. 55--61, Mar. 2020.

\bibitem{ITU:2017}
{ITU}, ``{Minimum requirements related to technical performance for {IMT-2020}
  radio interface(s)},'' \emph{ITU-R Standard M.2410-0}, Nov. 2017.

\bibitem{Jiang:21}
W.~Jiang, B.~Han, M.~A. Habibi, and H.~D. Schotten, ``The road towards {6G}: A
  comprehensive survey,'' \emph{IEEE Open J. Commun. Soc.}, vol.~2, pp.
  334--366, 2021.

\bibitem{Chae:16}
S.~H. Chae, S.-W. Jeon, and S.~H. Lim, ``Fundamental limits of spectrum sharing
  full-duplex multicell networks,'' \emph{{IEEE} J. Select. Areas Commun.},
  vol.~34, no.~11, pp. 3048--3061, Nov. 2016.

\bibitem{Chen:20}
S.~Chen, Y.-C. Liang, S.~Sun, S.~Kang, W.~Cheng, and M.~Peng, ``Vision,
  requirements, and technology trend of {6G}: {H}ow to tackle the challenges of
  system coverage, capacity, user data-rate and movement speed,'' \emph{IEEE
  Wirel. Commun.}, vol.~27, no.~2, pp. 218--228, Apr. 2020.

\bibitem{Zhang:19}
Z.~Zhang, Y.~Xiao, Z.~Ma, M.~Xiao, Z.~Ding, X.~Lei, G.~K. Karagiannidis, and
  P.~Fan, ``{6G} wireless networks: {V}ision, requirements, architecture, and
  key technologies,'' \emph{IEEE Veh. Technol. Mag.}, vol.~14, no.~3, pp.
  28--41, Sep. 2019.

\bibitem{Siddiqui:21}
M.~U.~A. Siddiqui, F.~Qamar, F.~Ahmed, Q.~N. Nguyen, and R.~Hassan,
  ``Interference management in {5G} and beyond network: {R}equirements,
  challenges and future directions,'' \emph{IEEE Access}, vol.~9, pp.
  68\,932--68\,965, 2021.

\bibitem{Saqib:21}
N.~U. Saqib, K.-Y. Cheon, S.~Park, and S.-W. Jeon, ``Joint optimization of {3D}
  hybrid beamforming and user scheduling for {2D} planar antenna systems,'' in
  \emph{Proc. {IEEE} {International Conference on Information Networking
  (ICOIN)}}, Jeju Island, Korea, Jan. 2021.

\bibitem{Tse_wireless}
D.~Tse and P.~Viswanath, \emph{Fundamentals of Wireless communication}.\hskip
  1em plus 0.5em minus 0.4em\relax Cambridge University Press, 2005.

\bibitem{Shafi:17}
M.~Shafi, A.~F. Molisch, P.~J. Smith, T.~Haustein, P.~Zhu, P.~De~Silva,
  F.~Tufvesson, A.~Benjebbour, and G.~Wunder, ``5{G}: {A} tutorial overview of
  standards, trials, challenges, deployment, and practice,'' \emph{IEEE J. Sel.
  Areas Commun.}, vol.~35, no.~6, pp. 1201--1221, Jun. 2017.

\bibitem{Nadeem:19}
Q.-U.-A. Nadeem, A.~Kammoun, and M.-S. Alouini, ``Elevation beamforming with
  full dimension mimo architectures in {5G} systems: {A} tutorial,'' \emph{IEEE
  Commun. Surv. Tutor.}, vol.~21, no.~4, pp. 3238--3273, 2019.

\bibitem{Costello:07}
D.~J. Costello and G.~D. Forney, ``Channel coding: {T}he road to channel
  capacity,'' \emph{Proc. IEEE}, vol.~95, no.~6, pp. 1150--1177, Jun. 2007.

\bibitem{Kumar:09}
K.~R. Kumar, G.~Caire, and A.~L. Moustakas, ``Asymptotic performance of linear
  receivers in {{MIMO}} fading channels,'' \emph{{IEEE} Trans. Inf. Theory},
  vol.~55, no.~10, pp. 4398--4418, Oct. 2009.

\bibitem{Jiang:11}
Y.~Jiang, M.~K. Varanasi, and J.~Li, ``Performance analysis of {ZF} and {MMSE}
  equalizers for {{MIMO}} systems: An in-depth study of the high {SNR}
  regime,'' \emph{{IEEE} Trans. Inf. Theory}, vol.~57, no.~4, pp. 2008--2026,
  Apr. 2011.

\bibitem{Damen:03}
M.~O. Damen, H.~E. Gamal, and G.~Caire, ``On maximum-likelihood detection and
  the search for the closest lattice point,'' \emph{{IEEE} Trans. Inf. Theory},
  vol.~49, no.~10, pp. 2389--2402, Oct. 2003.

\bibitem{Hassibi:05}
B.~Hassibi and H.~Vikalo, ``On the sphere-decoding algorithm {I}. expected
  complexity,'' \emph{{IEEE} Trans. Signal Processing}, vol.~53, no.~8, pp.
  2806--2818, Aug. 2005.

\bibitem{Jalden:05}
J.~Jalden and B.~Ottersten, ``On the complexity of sphere decoding in digital
  communications,'' \emph{{IEEE} Trans. Signal Processing}, vol.~53, no.~4, pp.
  1474--1484, Apr. 2005.

\bibitem{Guo:06}
Z.~Guo and P.~Nilsson, ``Algorithm and implementation of the {$K$}-best sphere
  decoding for {{MIMO}} detection,'' \emph{{IEEE} J. Select. Areas Commun.},
  vol.~24, no.~3, pp. 491--503, Mar. 2006.

\bibitem{Zhan:14}
J.~Zhan, B.~Nazer, U.~Erez, and M.~Gastpar, ``Integer-forcing linear
  receivers,'' \emph{{IEEE} Trans. Inf. Theory}, vol.~60, no.~12, pp.
  7661--7685, Dec. 2014.

\bibitem{Nazer:11}
B.~Nazer and M.~Gastpar, ``Compute-and-forward: Harnessing interference through
  structured codes,'' \emph{{IEEE} Trans. Inf. Theory}, vol.~57, no.~10, pp.
  6463--6486, Oct. 2011.

\bibitem{Ordentlich:15}
O.~Ordentlich and U.~Erez, ``Precoded integer-forcing universally achieves the
  {{MIMO}} capacity to within a constant gap,'' \emph{{IEEE} Trans. Inf.
  Theory}, vol.~61, no.~1, pp. 323--340, Jan. 2015.

\bibitem{Regev:21}
Y.~Regev and U.~Erez, ``Precise performance characterization of precoded
  integer forcing applied to two parallel channels,'' \emph{{IEEE} Trans.
  Wireless Commun.}, vol.~20, no.~12, pp. 7920--7931, Dec. 2021.

\bibitem{Ordentlich:conf_13}
O.~Ordentlich, U.~Erez, and B.~Nazer, ``Successive integer-forcing and its
  sum-rate optimality,'' in \emph{Proc. 51st Annual Allerton Conf. on Commun.,
  Control, and Comput.}, Monticello, IL, Oct. 2013, pp. 282--292.

\bibitem{He:18}
W.~{He}, B.~{Nazer}, and S.~{Shamai Shitz}, ``Uplink-downlink duality for
  integer-forcing,'' \emph{{IEEE} Trans. Inf. Theory}, vol.~64, no.~3, pp.
  1992--2011, Mar. 2018.

\bibitem{Silva:17}
D.~{Silva}, G.~{Pivaro}, G.~{Fraidenraich}, and B.~{Aazhang}, ``On
  integer-forcing precoding for the {Gaussian} {MIMO} broadcast channel,''
  \emph{{IEEE} Trans. Wireless Commun.}, vol.~16, no.~7, pp. 4476--4488, Jul.
  2017.

\bibitem{Ahn:19}
S.-K. Ahn and S.~H. {Chae}, ``Blind integer-forcing interference alignment for
  downlink cellular networks,'' \emph{{IEEE} Commun. Lett.}, vol.~23, no.~2,
  pp. 306--309, Feb. 2019.

\bibitem{Venturelli:20}
R.~B. Venturelli and D.~Silva, ``Optimization of integer-forcing precoding for
  multi-user {MIMO} downlink,'' \emph{IEEE Wirel. Commun.}, vol.~9, no.~11, pp.
  1860--1864, Nov. 2020.

\bibitem{Ntranos:13}
V.~{Ntranos}, V.~R. {Cadambe}, B.~{Nazer}, and G.~{Caire}, ``Integer-forcing
  interference alignment,'' in \emph{Proc. {IEEE} Int. Symp. Information Theory
  (ISIT)}, Istanbul, Turkey, Jul. 2013.

\bibitem{Hong:13}
S.~{Hong} and G.~{Caire}, ``Structured lattice codes for {$2\times2\times2$}
  {MIMO} interference channel,'' in \emph{Proc. {IEEE} Int. Symp. Information
  Theory (ISIT)}, Istanbul, Turkey, Jul. 2013.

\bibitem{Ordentlich:14}
O.~Ordentlich, U.~Erez, and B.~Nazer, ``The approximate sum capacity of the
  symmetric {G}aussian {$K$}-user interference channel,'' \emph{{IEEE} Trans.
  Inf. Theory}, vol.~60, no.~6, pp. 3450--3482, Jun. 2014.

\bibitem{Azimi-Abarghouyi:18}
S.~M. Azimi-Abarghouyi, M.~Hejazi, B.~Makki, M.~Nasiri-Kenari, and T.~Svensson,
  ``Integer-forcing message recovering in interference channels,'' \emph{IEEE
  Trans. Veh. Technol.}, vol.~67, no.~5, pp. 4124--4135, May 2018.

\bibitem{Abarghouyi:16}
S.~M. {Azimi-Abarghouyi}, M.~{Nasiri-Kenari}, B.~{Maham}, and M.~{Hejazi},
  ``Integer forcing-and-forward transceiver design for {MIMO} multipair two-way
  relaying,'' \emph{{IEEE} Trans. Veh. Technol.}, vol.~65, no.~11, pp.
  8865--8877, Nov. 2016.

\bibitem{Jiang:19}
H.~Jiang, J.~Zhao, L.~Shen, H.~Cheng, and G.~Liu, ``Joint integer-forcing
  precoder design for {MIMO} multiuser relay system,'' \emph{IEEE Access},
  vol.~7, pp. 81\,875--81\,882, 2019.

\bibitem{Hejazi:16}
M.~Hejazi, S.~M. Azimi-Abarghouyi, B.~Makki, M.~Nasiri-Kenari, and T.~Svensson,
  ``Robust successive compute-and-forward over multiuser multirelay networks,''
  \emph{{IEEE} Trans. Veh. Technol.}, vol.~65, no.~10, pp. 8112--8129, 2016.

\bibitem{Ahn-Chae-Kim-Kim:21}
S.-K. Ahn, S.~H. Chae, K.~T. Kim, and Y.-H. Kim, ``Successive cancellation
  integer forcing via practical binary codes,'' submitted to \emph{IEEE Trans.
  on Wireless Commun.}, 2021.

\bibitem{Chae-Jeon-Ahn:19}
S.~H. Chae, S.-W. Jeon, and S.-K. Ahn, ``Spatially modulated integer-forcing
  transceivers with practical binary codes,'' \emph{IEEE Trans. on Wireless
  Commun.}, vol.~18, no.~12, pp. 5542--5556, Dec. 2019.

\bibitem{Costa87}
M.~H.~M. Costa and A.~E. Gamal, ``The capacity region of the discrete
  memoryless interference channel with strong interference,'' \emph{{IEEE}
  Trans. Inf. Theory}, vol.~33, no.~5, pp. 710--711, Sep. 1987.

\bibitem{Carleial75}
A.~B. Carleial, ``A case where interference does not reduce capacity,''
  \emph{{IEEE} Trans. Inf. Theory}, vol.~21, no.~5, pp. 569--570, Sep. 1975.

\bibitem{Sato81}
H.~Sato, ``The capacity of the {G}aussian interference channel under strong
  interference,'' \emph{{IEEE} Trans. Inf. Theory}, vol.~27, no.~6, pp.
  786--788, Nov. 1981.

\bibitem{LNIT}
A.~{El Gamal} and Y.-H. Kim, \emph{Lecture Notes on Network Information
  Theory}, 2010.

\bibitem{Annapureddy:09}
V.~S. Annapureddy and V.~V. Veeravalli, ``Gaussian interference networks: {S}um
  capacity in the low-interference regime and new outer bounds on the capacity
  region,'' \emph{{IEEE} Trans. Inf. Theory}, vol.~55, no.~7, pp. 3032--3050,
  Jul. 2009.

\bibitem{Motahari09}
A.~S. Motahari and A.~K. Khandani, ``Capacity bounds for the {G}aussian
  interference channel,'' \emph{{IEEE} Trans. Inf. Theory}, vol.~55, no.~2, pp.
  620--643, Feb. 2009.

\bibitem{Shang09}
X.~Shang, G.~Kramer, and B.~Chen, ``A new outer bound and the
  noisy-interference sum-rate capacity for {G}aussian interference channels,''
  \emph{{IEEE} Trans. Inf. Theory}, vol.~55, no.~2, pp. 689--699, Feb. 2009.

\bibitem{Han:81}
T.~S. Han and K.~Kobayashi, ``A new achievable rate region for the interference
  channel,'' \emph{{IEEE} Trans. Inf. Theory}, vol.~27, no.~1, pp. 49--60, Jan.
  1981.

\bibitem{Zamir_book}
R.~Zamir, \emph{Lattice Coding for Signals and Networks}.\hskip 1em plus 0.5em
  minus 0.4em\relax Cambridge, U.K.: Cambridge Univ. Press, 2014.

\bibitem{Erez:04}
U.~{Erez} and R.~{Zamir}, ``Achieving $\frac{1}{2} \log (1+\text{SNR})$ on the
  {AWGN} channel with lattice encoding and decoding,'' \emph{{IEEE} Trans. Inf.
  Theory}, vol.~50, no.~10, pp. 2293--2314, Oct. 2004.

\bibitem{Bakoury:15}
I.~E. Bakoury and B.~Nazer, ``The impact of channel variation on
  integer-forcing receivers,'' in \emph{Proc. IEEE Int. Symp. Inf. Theory
  (ISIT)}, Hong Kong, Jun. 2015.

\bibitem{LLL}
A.~K. Lenstra, J.~H.~W.~Lenstra, and L.~Lovász, ``Factoring polynomials with
  rational coefficients,'' \emph{Math. Ann.}, vol. 261, no.~4, pp. 515--534,
  1982.

\bibitem{Sakzad:15}
A.~{Sakzad} and E.~{Viterbo}, ``Full diversity unitary precoded
  integer-forcing,'' \emph{{IEEE} Trans. Wireless Commun.}, vol.~14, no.~8, pp.
  4316--4327, Aug. 2015.

\bibitem{Seethaler:04}
D.~Seethaler, G.~Matz, and F.~Hlawatsch, ``An efficient {MMSE}-based
  demodulator for {{MIMO}} bit-interleaved coded modulation,'' in \emph{Proc.
  IEEE Global Telecommun. Conf., (GLOBECOM)}, Dallas, TX, Dec. 2004.

\bibitem{Liu:09}
T.-H. Liu, ``Some results for the fast {MMSE-SIC} detection in spatially
  multiplexed {MIMO} systems,'' \emph{{IEEE} Trans. Wireless Commun.}, vol.~8,
  no.~11, pp. 5443--5448, Nov. 2009.

\bibitem{Gan:09}
Y.~H. Gan, C.~Ling, and W.~H. Mow, ``Complex lattice reduction algorithm for
  low-complexity full-diversity {MIMO} detection,'' \emph{{IEEE} Trans. Signal
  Processing}, vol.~57, no.~7, pp. 2701--2710, Jul. 2009.

\bibitem{Keyhole02}
D.~Chizhik, G.~J. Foschini, M.~J. Gans, and R.~A. Valenzuela, ``Keyholes,
  correlations, and capacities of multielement transmit and receive antennas,''
  \emph{{IEEE} Trans. Wireless Commun.}, vol.~1, no.~2, pp. 361--368, Apr.
  2002.

\bibitem{Legnain:13}
R.~M. Legnain, R.~H.~M. Hafez, I.~D. Marsland, and A.~M. Legnain, ``A novel
  spatial modulation using {MIMO} spatial multiplexing,'' in \emph{Proc.
  International Conference on Communications, Signal Processing, and their
  Applications (ICCSPA)}, Sharjah, United Arab Emirates, Feb. 2013.

\bibitem{Sato;81}
H.~Sato, ``The capacity of the {G}aussian interference channel under strong
  interference,'' \emph{{IEEE} Trans. Inf. Theory}, vol. IT-27, pp. 786--788,
  Nov. 1981.

\end{thebibliography}

\end{document}